**The Achilles tendon enthesis rebuilds its mineralization front on reloading but retains a nanoscale imprint of unloading.**

M.L. Stammer[1], C. Camy[2,3], M. Frewein[1], I. Silva Barreto[1], C. Genovesio[4], M. Eckermann[5], A. Karimbana[1], K. Iliopoulos[1], R. Ranjan[1], N. Wittig[6], T. Fovet[7], T. Brioche[7], A. Chopard[7], M. Burghammer[5], S. Brasselet[1], H. Birkedal[6], M. Pithioux[2,3,8], S. Roffino[2]‡, T.A. Grünewald‡[1,*]

[1] Aix-Marseille Université, CNRS, Centrale Med, Institut Fresnel, Marseille, France

[2] Aix-Marseille Université, CNRS, ISM, Institut des Sciences du Movement, Marseille, France

[3] Aix-Marseille Université, APHM, CNRS, ISM, Mecabio Facility, Anatomy Laboratory, Timone, Marseille, France

[4] Aix-Marseille Université, Faculté de Pharmacie, Marseille, France

[5] European Synchrotron Radiation Facility, Grenoble, France

[6] Aarhus University, Department of Chemistry, Aarhus, Denmark

[7] PhyMedExp, Univ Montpellier, Inserm, CNRS, Montpellier, France

[8] Aix-Marseille Université, APHM, CNRS, ISM, Sainte-Marguerite Hospital, Institute for Locomotion, Marseille, France

‡ equally contributing
*Correspondence: tilman.grunewald@fresnel.fr

## Abstract

The enthesis is a graded fibrocartilaginous interface that transfers load between tendon and bone, yet the nanoscale mechanisms stabilizing its mineralization front remain unclear. Here, we combine multimodal 2D/3D X-ray imaging with nonlinear optical microscopy to map structural, crystalline and extracellular matrix organization across the murine Achilles tendon enthesis under unloading and reloading. Unloading reduces the tidemark-associated two-photon fluorescence (2PF) peak and is accompanied by diffuse mineralization into previously unmineralized fibrocartilage. This unloading-associated mineral exhibits increased apparent crystallite size, an enlarged *c*-axis lattice parameter, reduced crystalline texture and a diminished collagen order gradient, consistent with an altered mineralization environment. Upon reloading, the 2PF peak recovers, but a new tidemark forms ~20 µm from the original boundary, creating a zone with a persistent nanoscale imprint in the mineral tessellation. These findings establish the enthesis as a mechanically governed graded interface in which matrix-mediated boundary control constrains mineral formation and in which a record of mechanical history is imprinted into the nanostructure.

**Introduction**

Graded interfaces are a core materials design principle: by distributing elastic mismatch across the gradient, they enable stress transfer between mechanically dissimilar constituents while suppressing interfacial stress concentrations[1–7]. In synthetic and bio-inspired materials, such gradients are usually fixed after fabrication[8,9]. In living tissues, however, graded interfaces must be maintained under continuously changing mechanical[10] and biochemical conditions[11], raising the question of how their boundaries remain stable over time.
The tendon-to-bone interface, the enthesis, is a demanding biological example of such a graded interface. It anchors the compliant tendon to the stiff bone through a fibrocartilaginous transition zone composed of unmineralized (UFc) and mineralized fibrocartilage (MFc)[2]. This prominent tissue structure is shared by several clinically important entheses including the Achilles tendon (Fig. 1a), supraspinatus and patellar tendon entheses, highlighting its fundamental role in musculoskeletal load transfer. However, the native graded fibrocartilaginous interface is not reliably restored after rupture, avulsion or surgical repair, often resulting in persistent functional impairment[12,13]. The broader clinical relevance of tendon-to-bone interfaces is underscored by rotator cuff repair, an enthesis-associated injury, for which more than 250,000 procedures are performed annually in the United States, with estimated societal costs exceeding US$3.4 billion[14].
The murine Achilles tendon enthesis (Fig. 1b) is hierarchically organized across molecular to tissue-level length scales with a gradient in mechanical properties. The main components of the fibrocartilaginous transition zone are type I and type II collagens embedded in a non-collagenous matrix, rich in proteoglycans and other non-collagenous proteins. In the MFc, the organic matrix additionally contains bioapatite mineral particles (Fig. 1c). At the mineralization front, histologically referred to as the tidemark, mineral deposition in a collagen-rich scaffold establishes a sharp, yet compliant transition between unmineralized and mineralized fibrocartilage[1,3,11]. This boundary must be maintained under continuous mechanical stimulation.

Mechanical loading is known to regulate the structure and composition of fibrocartilaginous interfaces[15–19]. Analogous loading-dependent remodeling of collagen nanostructure and mechanical competence has been demonstrated in the tendon itself[20,21]. In the enthesis, unloading alters the morphology, nanoscale mineral crystallography and mechanical behavior[17,22]. At the micrometer scale, the position of the mineralization front is also known to be load-sensitive as reloading after a period of unloading induces the formation of a new tidemark at the enthesis[17]. Likewise, in rat articular cartilage, unloading nearly doubles the tidemark mineral apposition rate[23] . Whether this reflects a nanoscale change in the mineral itself, or the displacement of an otherwise unchanged boundary, is unknown.

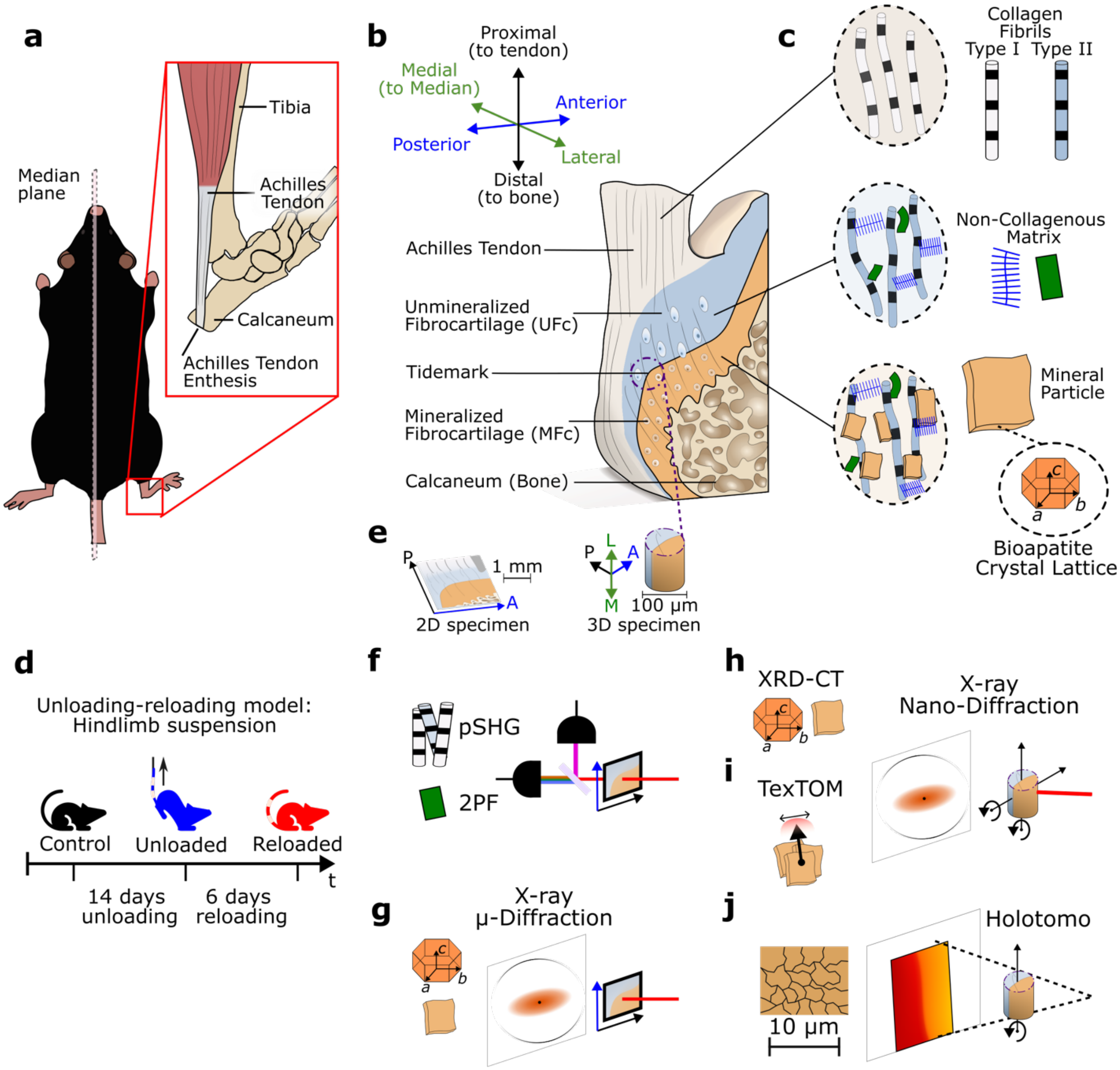

*Figure 1: | **Multimodal mapping strategy for loading-induced remodeling at the Achilles tendon enthesis.** **a** Murine hindlimb anatomy and location of the Achilles tendon enthesis at the calcaneal insertion. **b** Schematic sagittal view of the right Achilles tendon enthesis, showing the Achilles tendon, unmineralized fibrocartilage (UFc), mineralized fibrocartilage (MFc), calcaneal bone and the tidemark separating UFc from MFc. The sagittal section plane (blue dotted line) shown was used for the two-dimensional analyses (e-f) and 3D interface-pillar volume (purple) from the tidemark was used for the 3D measurements (g-i) **c** The principal nanoscale constituents of the Achilles tendon enthesis fibrocartilage: collagen fibrils (Collagen type I and II), the non-collagenous matrix, rich in proteoglycans (blue symbols) and non-collagenous proteins (green symbols), and the bioapatite mineral particles and their crystal lattice. **d** Loading scheme design. Control samples were compared to a 14-day hindlimb unloading tail-suspension group, followed by 6-day reloading group. **e** Specimen geometry. 2D sagittal sections were prepared for large field-of-view measurements. 3D specimen enabled nano-beam X-ray imaging. **f** Section-based pSHG and 2PF quantify collagen organization and non-collagenous matrix (2PF) intensity **g** Scanning X-ray microdiffraction determines scattering derived mineral parameters (SAXS) and crystalline properties (WAXS) **h** XRD-CT on the 3D sample pillar maps the crystalline properties **i** Texture tomography on the 3D pillar determines the mineral texture **j** Holotomography gives quantitative mass-density contrast and high-resolution morphology information*

Here we combine multimodal two- and three-dimensional X-ray imaging with nonlinear optical microscopy to map murine Achilles tendon enthesis after hindlimb unloading and subsequent reloading (Fig. 1d).
Polarization-resolved second-harmonic generation and two-photon fluorescence resolve collagen organization and non-collagenous matrix (2PF) intensity in thin sections (Fig. 1f), while scanning X-ray microdiffraction (Fig. 1g), X-ray diffraction tomography (XRD-CT)[24,25] (Fig. 1h), texture tomography (TexTOM)[26] (Fig. 1i) and holotomography (Fig. 1j) quantify mineral crystallography (scanning X-ray microdiffraction and XRD-CT), texture (TexTOM) and quantitative mass density contrast (holotomography) in two and three dimensions.

**Results**

To determine how altered mechanical loading affects the nanoscale organization of the Achilles tendon enthesis, we compared control samples with entheses subjected to 14 days of hindlimb unloading and entheses reloaded for 6 days after unloading.

**Unloading depletes the tidemark 2PF peak and flattens the collagen order gradient**

To understand the mechanism that governs the mineralization front, we quantified changes in the amount of extracellular matrix (ECM) components and collagen organization across the tidemark using combined two-photon fluorescence (2PF)[27,28] and polarization-resolved second harmonic generation (pSHG)[29]. While the 2PF intensity is non-specific and detects a broad range of non-collagenous proteins such as elastin[30] and collagen cross links[28], it has little sensitivity to proteoglycans[31]. Together, these measurements report on non-collagenous matrix proteins and collagen order in the local environment of the enthesis.

In control entheses, 2PF reveals a distinct enrichment of non-collagenous matrix proteins near the tidemark within the MFc (Fig. 2a, d), while pSHG shows higher collagen orientational order in the mineralized fibrocartilage (Fig. 2 f , i) and a graded decrease across ~50 µm into the UFc (Fig. 2i, j). After 14 days of unloading, the tidemark-associated 2PF peak is strongly reduced (Fig. 2d - e) and the compositional contrast across the interface is reduced (Fig. 2b). This is accompanied by a proteoglycan depletion in the UFc as shown by toluidine blue staining (SI Fig. 1h). The collagen order profile gradually decreases across the tidemark into the UFc in the control, this decrease happens over a shorter distance and with reduced amplitude in unloading (Fig. 2i, j). After 6 days of reloading, the 2PF peak reappears and aligns with the re-established interface (Fig. 2c), collagen order partially recovers toward control-like MFc/UFc values (Fig. 2i, j) and proteoglycan content in the UFc recovers (SI Fig. 1i).

Across all groups, collagen anisotropy is consistently higher in the MFc than in the UFc, consistent with a mineral-imposed ordering reported previously[36]. Additionally, the total SHG signal is lower in the MFc (SI Fig. 2), indicating a mineralization-dependent density change of fibrillar collagen[29]. The largest loading-dependent changes in collagen order were confined to the UFc, whereas MFc values remained comparatively stable over the unloading-reloading protocol.

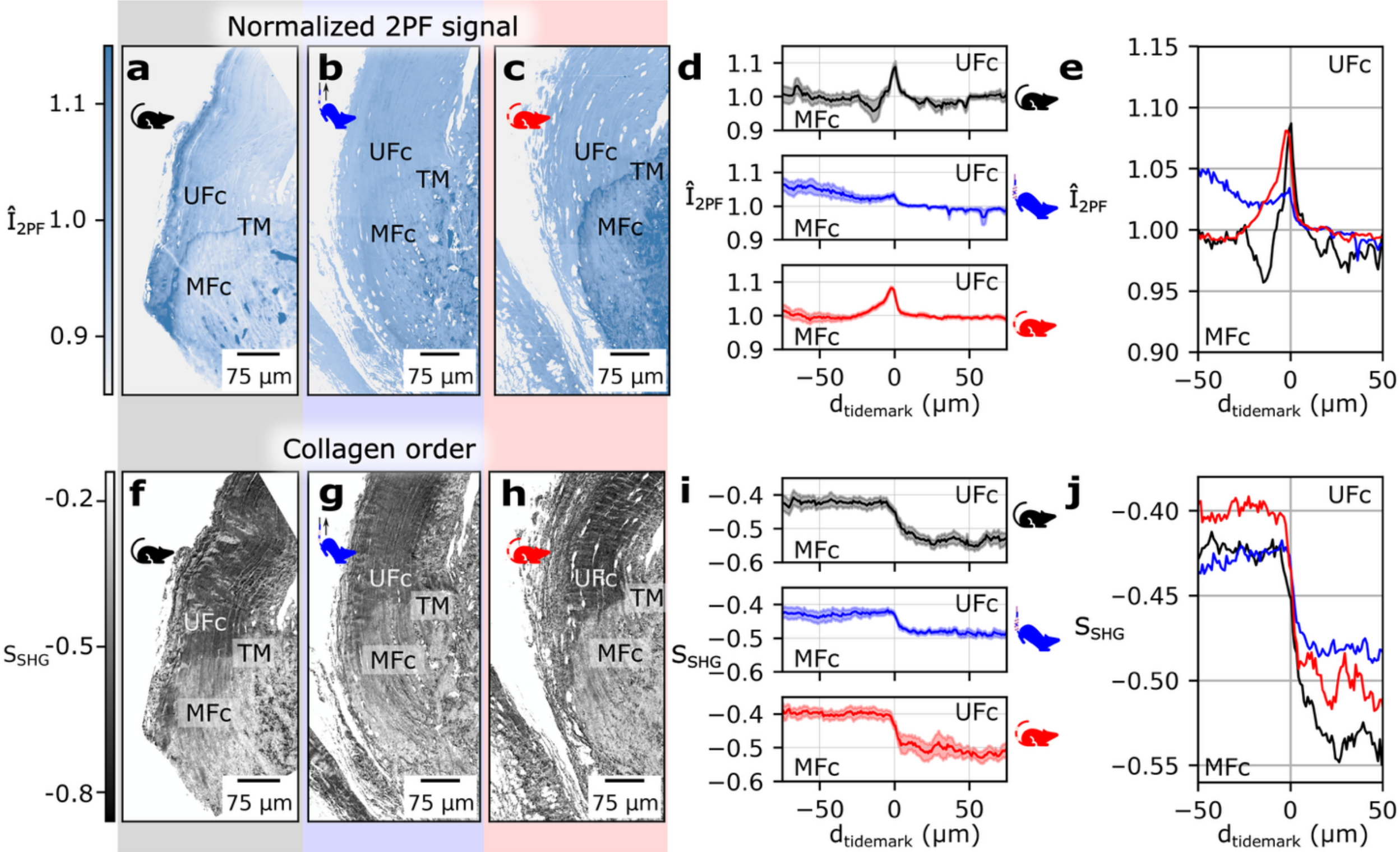

*Figure 2:* ***Load-dependent changes in the tidemark-associated non-collagenous matrix (2PF) peak and collagen order*** *.* ***a-c*** *Two-photon fluorescence (2PF) maps of control (CTL, black mouse), unloading (UL, blue mouse) and reloading (RL, red mouse) enthesis reveal a tidemark-associated 2PF peak in CTL and RL that is strongly reduced after unloading.* ***d*** *Group-averaged 2PF profiles plotted against Euclidean distance to the tidemark (N = 5, 5 and 3 for CTL, UL and RL, respectively, standard error estimates population sampling)* ***e*** *Overlay of group averaged 2PF indicates partial recovery of the 2PF peak in the MFc near the tidemark.* ***f-h****, pSHG-derived collagen order parameter maps for CTL, UL and RL. $S_{SHG}$ values decrease with a decrease of molecular order.* ***i, j*** *Distance-resolved collagen order shows a gradual decrease from MFc to UFc in CTL, compression of this gradient after UL and partial recovery for RL. Together, 2PF and pSHG indicate that both the 2PF peak and collagen order gradients at the tidemark are loading dependent.*

## Mineral formed during unloading is crystallographically distinct

To determine whether altered mechanical loading leaves a three-dimensional nanostructural imprint in mineral crystallography and texture, we performed XRD-CT (Fig. 1h) and TexTOM (Fig. 1i) on ~100 µm pillars centered on the tidemark region of the Achilles tendon enthesis (Fig. 1b) for a sample of each group. This region was selected because histology of the same model predicts the largest loading-induced difference[17,19]. XRD-CT and voxel-wise Rietveld-refinement profile fitting provides voxel-resolved maps of mineral crystallographic parameters like *c*-axis lattice parameter (Fig. 3a, b) or the SAXS-derived *T* parameter as a particle-thickness/nanostructural proxy (Fig. 3c, d). TexTOM reconstructs, from the same set of diffraction data used for XRD-CT, the three-dimensional crystallite orientation distribution function (ODF) [37], from which we extract the *c*-axis nematic director and order parameter to determine the *c*-axis orientation information (Fig. 3e - g). Because unloading and reloading alters the position of the tidemark, we distinguish between the original tidemark, corresponding to the native MFc boundary, and the newly formed tidemark re-established after reloading. Unless stated otherwise, distance-resolved profiles are referenced to the original tidemark in CTL and UL, and to the new tidemark in RL. (SI Fig. 3 and 4).

In control entheses, the mineral *c*-axis showed a coherent preferred orientation within the mineralized fibrocartilage, with local perturbations around chondrocyte lacunae (SI Fig. 5). Mineral crystallographic parameters and nematic order varied gradually across the tidemark, defining a characteristic gradient that serves as the reference against which unloading- and reloading-induced changes are assessed. In the unloaded specimen, a 20–30 µm diffuse mineralization zone (DM) appeared at positive distances from the tidemark, within what was previously unmineralized fibrocartilage. This region showed an increased *T*-parameter, an expanded *c*-axis and reduced texture order and average local misalignment compared to the remaining MFc (Fig. 3a, c). Even if the *T*-parameter depends on both mineral particle dimensions and the degree of mineralization, its increase demonstrates a modified mineral nanostructure rather than a simple extension of native MFC. Collectively, these changes demonstrate that mineral particles formed during unloading differ from native MFc in their nanoscale organization.

Reloading restores bulk MFc parameters towards control-like values but the region between the original tidemark and the newly formed tidemark remain distinct. This region shows spatial variations in mineral properties (Fig. 3 b, d) and texture (Fig. 3 f, g), different from the native MFc structure found in the control group. Thus, short-term reloading partially restores native MFc mineral properties while preserving a nanoscale record of the unloading-affected mineralization zone.

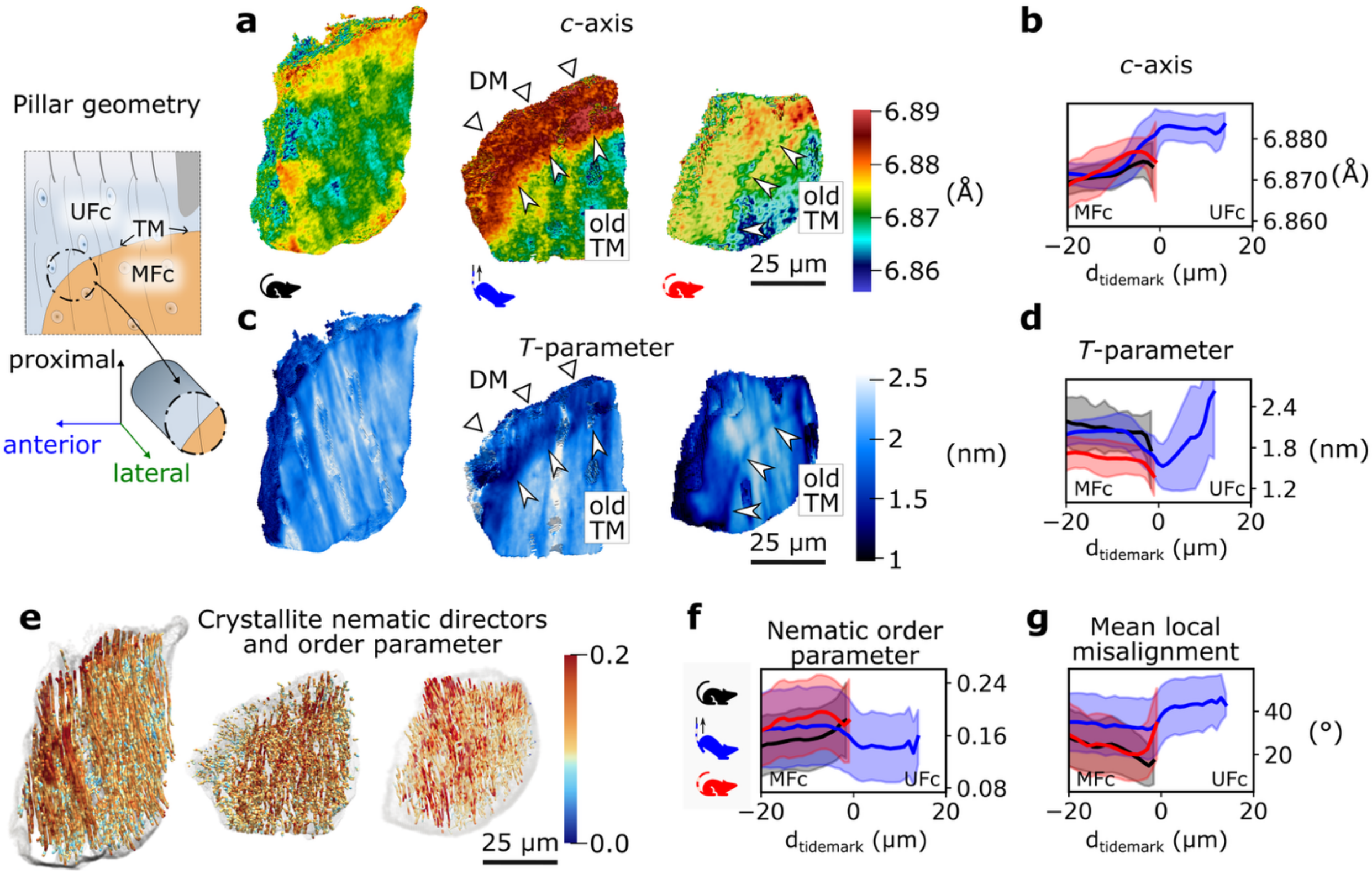

*Figure 3: **Crystalline texture remodeling at the tidemark and persistent localized heterogeneity in mineral properties and texture with modified loading**. **a** 3D maps of XRD-CT reconstructed c lattice parameter in CTL, UL and RL (narrow arrowheads mark the original TM in UL and RL, triangular arrowheads mark the diffuse mineralization region). The sketch indicates the sampling geometry across the tidemark region **b** Distance-resolved c-lattice parameter profiles extracted from the 3D volumes for CTL (black), UL (blue) and RL (red) **c** 3D maps of XRD-CT reconstructed T parameter in CTL, UL and RL **d**. Distance-resolved T parameter profiles extracted from the 3D volumes for CTL (black), UL (blue) and RL (red) **e** 3D maps of TexTOM reconstructed crystallite nematic directors and orders parameters, visualized as an orientation flow field. **f** Distance-resolved nematic order parameter (mean ± SD) **g** mean local misalignment. Distances are referenced to the relevant tidemark for each condition: the original tidemark for CTL and UL and the newly formed tidemark for RL. In UL and RL the original tidemark is still visible and marked with white arrowheads.*

## Mineral changes at the interface reproduce across CTL, UL and RL groups

To assess whether the mineral heterogeneity identified in the 3D pillar specimens was reproducible at the population level, we performed scanning X-ray microdiffraction (µXRD) on sagittal enthesis sections from $N$ = 5, 6, and 7 specimens obtained from 4, 4, and 5 animals in the CTL, UL, and RL groups, respectively. Pixel-wise Rietveld refinement and SAXS fitting yielded maps of apparent crystallite size along the *c*-axis (ACS), the *T*-parameter, a scale-factor mineral-density proxy, and *c*-axis lattice parameter. Mineral values fall within the expected range for murine bioapatite [38,39]. In CTL, mineral parameters varied only moderately within tidemark-proximal MFc, consistent with the 3D data. Unloading produced a ~20 µm DM zone (triangular arrowheads, Fig. 4b, g) characterized by elevated *T*-parameter and ACS, an extended *c*-axis plateau, and reduced mineral density relative to native MFc (Fig. 4d, e, i, j). After reloading, bulk MFc parameters shifted back toward CTL levels, but a discontinuity persisted within the mineralized tissue at the position of the newly formed tidemark (white arrowheads, Fig. 4c, h). These observations reproduce the persistent heterogeneity observed in the 3D specimens at population scale.

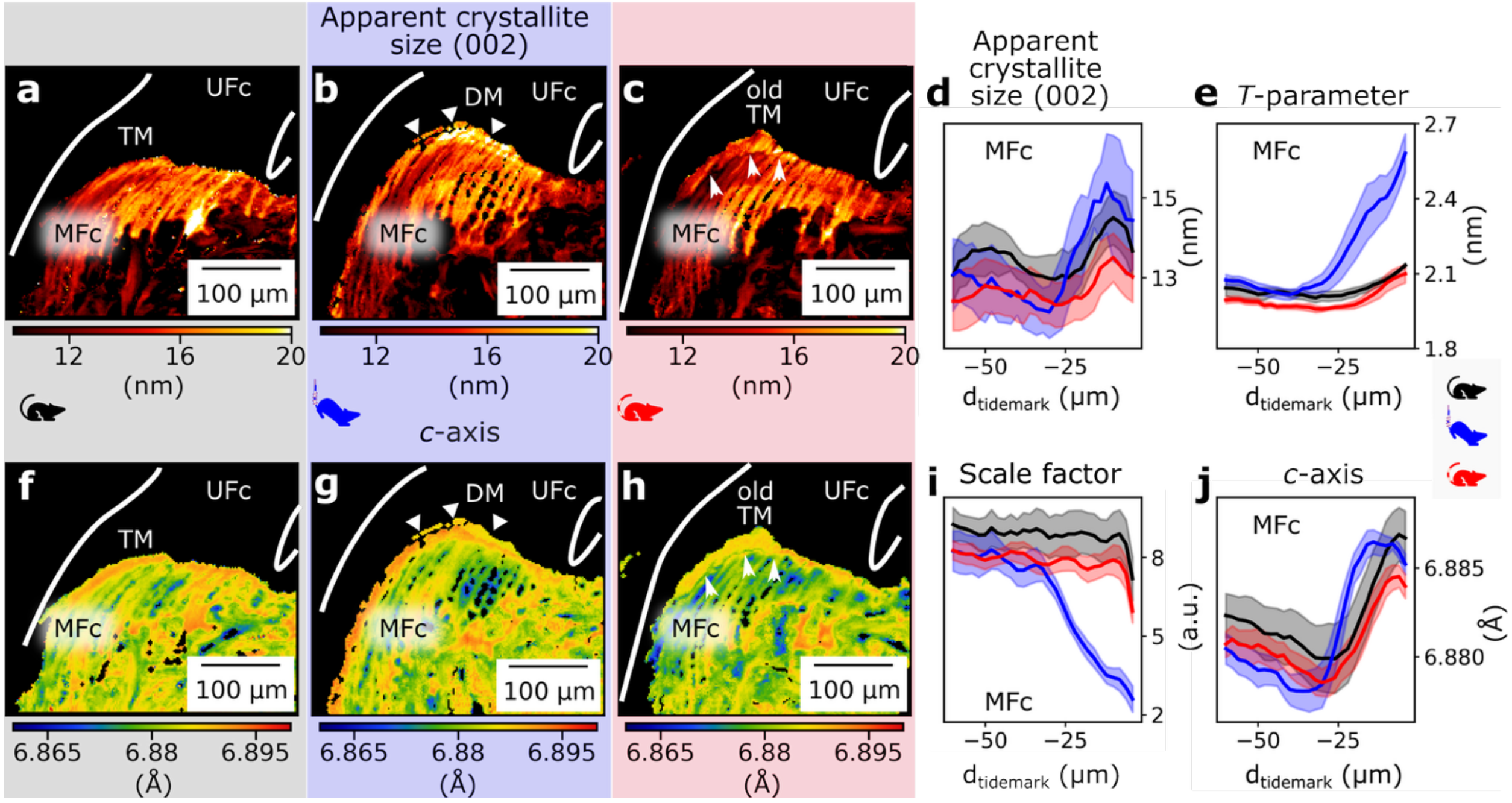

*Figure 4: **Distance-resolved scanning X-ray microdiffraction confirms localized mineral changes from loading changes**. **a-c** 2D-µD maps of apparent crystallite size along c-axis lattice direction (ACS) and **f-h** 2D-µD maps of c-axis lattice parameter [DM = diffuse mineralization zone, triangular arrowheads] [native TM, narrow arrowheads] **d,e,i,j** Group-averaged mineral parameters (ACS, T-parameter, scale factor, and c-axis lattice parameter, respectively) as a function of distance from the tidemark (mean ± s.e.m.).*

*Unloading broadens the mineralized region by forming the DM zone with elevated T-parameter and ACS and reduced mineral density; reloading restores bulk MFc values while leaving a discontinuity at the newly formed tidemark position.*

## Reloading rebuilds a sharp boundary without restoring the native mineral architecture

To assess how loading-induced mineral changes manifest in the three-dimensional morphology of the mineralized interface, we used holotomography to map mass-density contrast across the tidemark region.

In control entheses, the mineralized fibrocartilage formed a coherent, sharply delineated domain adjacent to the tidemark with no detectable diffuse mineralization zone extending into the UFc (Fig. 5a). After unloading, mineral extended beyond the original tidemark into the previously unmineralized zone, forming a diffuse, unbounded zone of newly formed MFc (DM) (Fig. 5b). After reloading, a more sharply defined MFc domain was visible beyond the original tidemark, consistent with the reconstruction of the mineralization boundary at a new position (Fig. 5c).

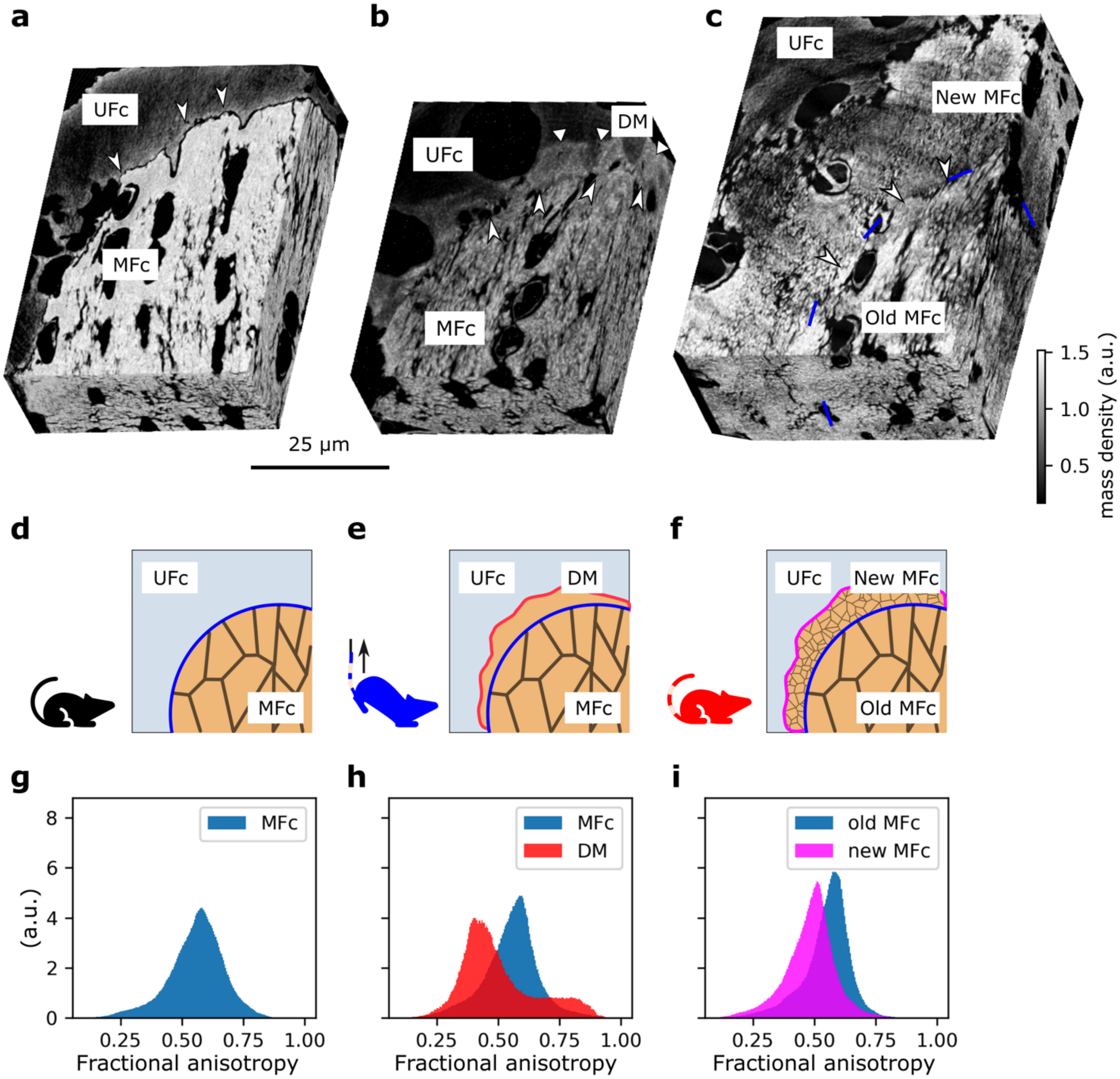

*Figure 5:* ***Holotomography resolves loading-dependent changes in mineralized fibrocartilage morphology and tessellation. a–c*** *Mass-density volumes from holotomographic reconstruction showing the tidemark region in CTL, UL, and RL. Native MFc forms a coherent, sharply bounded domain in CTL; unloading produces the diffuse mineralization zone (DM) extending beyond the original tidemark; reloading re-establishes a more sharply defined boundary at a new position (narrow arrowheads demark the original tidemark, triangular arrowheads the DM zone;* ***c:*** *original tidemark outlined in blue).* ***d–f*** *Sketched changes across the specimens illustrating the formation of DM (red) beyond the original tidemark (blue). After reloading the newly formed MFc (magenta) is visibly textured, however distinct from the native MFc.* ***g–i*** *Distributions of local fractional anisotropy (FA) within each volume:* ***g****, CTL MFc;* ***h****, UL native MFc vs DM;* ***i****, RL native vs newly formed MFc. Distributions are area-normalized; the sampling unit is a structure-tensor window within one specimen, so the panel describe within-specimen architecture are not used for between-group inference*

Within each specimen, mineral formed beyond the original tidemark was less dense than the native MFc in the same volume by ~ -0.24 for the DM in UL and ~ -0.15 for the newly formed MFc in RL (SI Fig. 7, SI Table 1). The densities extracted here should not be treated as quantitative values due to per-specimen offsets in the phase-retrieval procedure, we compare only differences within each specimen.
We used the same volumes to assess whether the newly formed mineral reproduces the mesoscale tessellation of the native MFc - the partitioning of the MFc into mineral domains separated by lower-density organic interfaces[40].
Tessellation produces locally parallel density interfaces, which we quantified as the fractional anisotropy (FA) of the local structure tensor: high where the interfaces are locally aligned, low where the organization is isotropic (SI Fig. 6). Control MFc gave a single, unimodal FA distribution (0.56 ± 0.11; Fig. 5g) In RL, the newly formed MFc distribution was shifted uniformly to lower FA relative to the native MFc of the same volume (0.45 ± 0.08 vs 0.53 ± 0.08; Fig. 5i), indicating a less regularly tessellated architecture. In UL the DM and the adjacent native MFc had similar mean FA (0.50 ± 0.13 vs 0.49 ± 0.09), but the DM distribution was twice as broad in variance and carried a secondary high-FA population (Fig. 5h). The mean is therefore uninformative for the DM; which is better described as architecturally heterogenous than as uniformly disordered: it contains both mineral with weaker interface alignment than native MFc and regions with stronger alignment than any part of it. The actual statistical significance levels are reported in SI Table 1.
Given the specimen-level sample sizes for the 3D data (SI Table 1), this FA difference should be interpreted as a consistent directional trend corroborated by the population-level µXRD data, rather than a statistically powered standalone result.

**Discussion**

The main finding of this study is that the enthesis tidemark is not merely a histological landmark, but a mechanically maintained boundary condition for mineral growth. The tendon-to-bone attachment has long been understood as a functionally graded interface in which collagen and mineral gradients distribute mechanical mismatch across the transition from tendon to bone[1,3,4]. Our data extend this view by showing that the organization of this gradient itself is load-sensitive, postulating a pressure-mediated inhibition. In the murine Achilles tendon enthesis, the tidemark is not only defined by the beginning of mineralized tissue but by the impediment of further mineralization. The relevant mechanical design problem is not only how a mineralization gradient transfers load, but how the boundary of this gradient is actively maintained.
That reduced loading permits the mineralization front to advance is not new. Carter and Wong proposed that cyclic hydrostatic pressure during weight bearing inhibits advancement of the subchondral mineralization front[41], consistent with O'Connor's observation that hindlimb unloading nearly doubles the tidemark apposition rate in the rat knee (2.17 ± 0.86 µm/day vs 1.12 ± 0.12 µm/day), thickening the calcified layer at the expense of the uncalcified one[23]. Applied to our 14-day unloading period, that rate predicts ~30 µm of advance - consistent with the 20 - 30 µm diffuse mineralization zone, despite differences in species and tissue. Two features of our data are not accounted for by pressure-mediated inhibition alone. First, the effect occurs here at a mature, quiescent front rather than the actively advancing front of a growing animal, the case O'Connor identified as unresolved. Second, and more fundamentally, a purely mechanical inhibition only predicts that the front advances, not that it deposits different material: mineral formed during unloading differs from native MFc in *c*-axis lattice parameter, *T*-parameter, crystalline texture, tessellation and mass density. Advance and altered architecture are therefore separable, and the latter points to a change in the local matrix environment in which mineral forms.

Reloading shows that recovery is not equivalent to reversal. Previous work in the same murine Achilles tendon unloading-reloading model reported structural and mechanical changes, including the appearance of a new tidemark after reloading[17,19]. This study adds that the appearance of this new tidemark is not just a histological repositioning but is accompanied by persistent nanoscale and mesoscale mineral differences. The enthesis re-establishes a sharp boundary, but does so at a new position within the previously unmineralized fibrocartilage. The region between the original and newly formed tidemarks therefore becomes a structural record of the unloading episode[10]. The interface recovers the boundary sharpness at the tissue scale, while retaining altered crystallographic, textural and tessellation features within the mineralized volume.

Two spatially distinct, load-sensitive matrix compartments accompany this behavior: a non-collagenous matrix (2PF) peak in the MFc next to the tidemark (Fig. 2), and a proteoglycan depletion in the adjacent UFc (toluidine blue, SI Fig. 1g - i). Histochemistry of the tidemark indicates that it is collagen- and glycoprotein-rich but lacks the glycosaminoglycans of conventional proteoglycans[42], so the tidemark peak and the UFc proteoglycans are unlikely to report the same structure. A candidate for the first is available in this tissue: osteopontin, a non-collagenous mineralization-inhibiting phosphoprotein, is concentrated at the UFc-MFc mineralization front of the wild-type murine Achilles tendon enthesis[40]. Under the stenciling principle, extracellular matrix mineralization is patterned by local removal of such inhibitors rather than the delivery of promoters[43]. This framework accommodates our observation without requiring inhibitor abundance to change: in Hyp mice, osteopontin is present but delocalized, and the mineralization front is correspondingly ill-defined with defective tessellation[40]. In the genetic and mechanical case alike, it is the spatial patterning of the inhibition rather than its magnitude that varies the boundary sharpness and tessellation coherence. But it is important to keep in mind that several inhibition pathways based on phosphoproteins, pyrophosphates and e.g. citrates exist and most likely are activate at the same time[44–46].
The proteoglycan-rich UFc may contribute separately, by setting hydration, ion mobility and interfibrillar spacing[47]; reduced loading at this insertion is accompanied by muscle atrophy and altered fibrocartilage matrix composition[22,48–50]. The collagen-rich UFc is not inert with respect to mineralization as constrained local matrix constraints can lead to bioapatite formation in this hydrated scaffold[43,51,52]. Local collagen fibril orientation guides mineral growth direction and morphology across mineralizing tissue including enthesis fibrocartillage[53], providing a structural basis by which the loss of collagen order gradient observed here would alter the crystallographic texture and properties of the newly formed mineral. The present data do not identify the responsible molecular species but they define two load-sensitive matrix compartments that co-vary with the sharpness of the boundary and the architecture of the mineral formed beyond it.
This persistent difference in tessellation is consistent with altered mineralization conditions during unloading: although mineral deposition resumes upon reloading and scattering-derived properties recover toward control levels, the mineralization environment appears to impose longer-lived constraints on subsequent mineral growth.

The mineral formed after unloading appears to bear an imprint of the boundary conditions under which it was developed. Altered lattice parameters, SAXS-derived *T*-parameter, crystallite size, crystalline texture, mass density and tessellation, changes qualitatively consistent with prior reports of unloading-induced crystallographic disruption at the enthesis[22], indicate that this is not a mere extension of the MFc mineral into a new location. Because the *T*-parameter depends on both mineral particle size and degree of mineralization, we interpret it here as a nanostructural proxy rather than a direct particle-size measurement. Rather, these combined changes suggest mineral formation under altered protein matrix conditions. This interpretation is consistent with the idea that mineral architecture is shaped by the local tissue context, as seen in the enthesis mineral tessellation[40] and heterotopic mineralization of Achilles tendon, where ectopic deposits acquire a tissue-specific internal architecture rather than behaving as generic bone-like mineral[54]. This distinction between re-establishing the boundary and actual nano- and mesoscale architectural restoration may be important for understanding why short-term remobilization can improve some aspects of enthesis organization while leaving persistent structural defects[55].

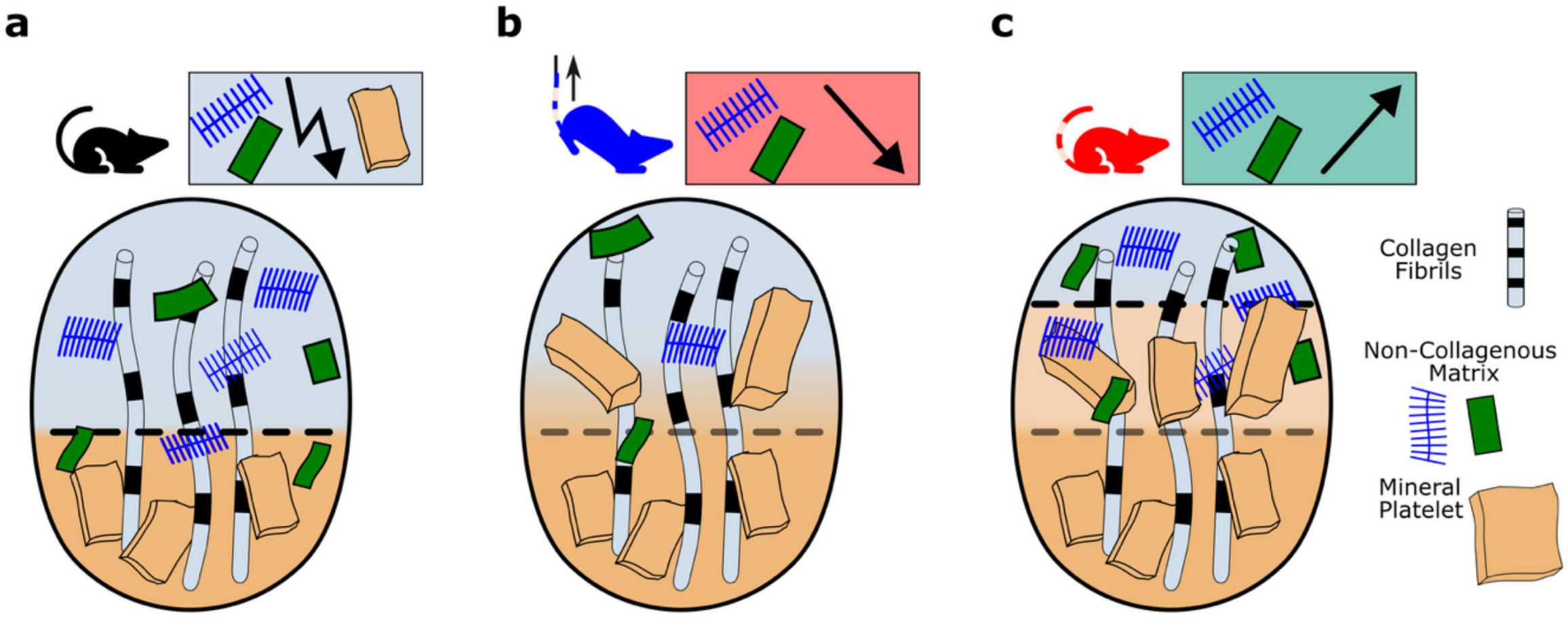

*Figure 6:* ***Sketch of load-dependent patterning of the mineralization boundary a*** *Under physiological load, a tidemark-associated non-collagenous matrix compartment coincides with a sharp boundary and coherent mineral tessellation* ***b*** *Unloading depletes this compartment; mineral forms diffusely beyond the original tidemark with altered crystallography and tessellation* ***c*** *Reloading re-establishes a sharp boundary at the new position, leaving the intervening zone structurally distinct.*

Taken together, these observations support the mechano-chemical model sketched in Fig. 6. Under physiological load, a tidemark-associated non-collagenous matrix compartment coincides with a sharp mineralization boundary and coherent mineral tessellation. Unloading depletes this compartment, and mineral forms diffusely beyond the original tidemark with altered crystallography, texture and tessellation. Reloading restores the matrix compartment and a confined boundary, but at a new position, leaving the intervening zone structurally distinct. This model is an interpretation consistent with our observations rather than a demonstrated causal chain: we measure co-variation between a matrix signal and mineral architecture, not the action of a specific molecule on mineral growth.
The principal limitation of this study is that the molecular identity of the tidemark-associated matrix compartment remains unresolved. Label-free 2PF and toluidine-blue staining identify two load-dependent matrix compartments, but they do not distinguish specific proteoglycans, non-collagenous proteins or changes in molecular state.
Resolving the molecular identity of these compartments will require combining targeted immunolabelling, spatial proteomics or genetic perturbation with multimodal X-ray imaging could help in identifying the components of these compartments and investigate how they regulate collagen order, mineral density and mineral growth. Clinically, these findings suggest that strategies for enthesis repair and biomimetic design[12,13,56] must account for the dynamic regulation of the mineralization boundary, not only the static compositional gradient. Restoring matrix-mediated boundary control under load may therefore be as critical a design requirement as recreating the collagen-to-mineral transition itself.

## Materials and Methods

### Animals and experimental design

Eighteen 12-week-old male C57BL/6J mice were used in this study. All procedures were approved by the Ethics Committee for Animal Experimentation of Languedoc-Roussillon and complied with the French National Research Council guidelines for the care and use of laboratory animals (APAFIS#28764-2020122115407491). Mice were housed individually at 22 °C under a 12 h/12 h light–dark cycle with ad libitum access to standard chow and water, and all efforts were made to minimize suffering.

Mice were assigned to three groups of six. Control animals (CTL) were maintained under standard housing. The remaining twelve underwent 14 days of hindlimb unloading by tail suspension following Morey-Holton and Globus[57]; six were sacrificed immediately after unloading (UL) and six were allowed 6 days of free reambulation before sacrifice (RL). After sacrifice, the right ankle of each animal was dissected, fixed in 4% paraformaldehyde, dehydrated through a graded ethanol series, cleared in methylcyclohexane, and embedded in methyl methacrylate (MMA).

### Two-dimensional specimens

Sagittal sections through the median region of the Achilles tendon enthesis of 5 µm nominal thickness were cut from MMA-embedded samples using a microtome (Leica RM2265). For scanning X-ray microdiffraction, sections were mounted on SiNx membranes (Silson, 4mm window size, 1 µm thickness, 7.5mm frame size). For polarization-resolved SHG, two-photon fluorescence and proteoglycan staining, sections were mounted on Superfrost Plus microscopy slides. For proteoglycan staining, sections were stained with toluidine blue (pH 5), dehydrated, and mounted with Entellan.

### Three-dimensional specimens - pillar preparation

Approximately 1 mm³ tissue blocks were excised from MMA-embedded ankles using a diamond-blade precision saw (Buehler IsoMet High Speed). Regions of interest (~100 µm across) containing both unmineralized and mineralized fibrocartilage spanning the tidemark were selected by laboratory micro-computed tomography (Zeiss Xradia 620 Versa; 40 kV; 4 µm voxel size), following the process of Wittig et al. [58]. Blocks were mounted face-down with two-component epoxy on a custom MMA sample holder, aligned so that the center of the region of interest coincided with the rotation axis of the holder, and verified by a second laboratory tomogram at 1 µm voxel size. Coarse pillars of 100–200 µm diameter were produced by mechanical micro-milling following Holler et al[59] as described in Ref[58] (see SI Fig. 8 for a detailed description of the workflow).

Final pillar shaping was performed by one of two methods. Samples CTL, UL, and RL_1 were shaped by xenon plasma focused ion beam milling (PFIB; Tescan Amber). A thin conductive carbon layer was deposited prior to milling to suppress surface charging. Milling currents ranged from 0.1 to 1 µA under continuous SEM monitoring of milling progress.
Sample RL_2 was shaped by femtosecond laser ablation (Yuja, Amplitude Systems, hybrid (crystal/fiber), passively mode-locked laser, λ = 1030 nm; pulse duration 400 fs; 2.5–5 µJ per pulse, adjusted to local material density) using an Olympus LMPLFLN-BD 20× objective (NA = 0.4), producing a focal spot of ~3 µm diameter and a Rayleigh length of ~10 µm. Single-shot ablation was used at each position to minimize thermal damage, with an axial step size of 6 µm; lateral step size and pulse energy were adapted to the local geometry of the mineralized fibrocartilage. Ablation parameters were optimized on calcaneal tissue excluded from the region of interest during coarse milling. Scanning geometry was calibrated by patterning a fiducial structure into the sample block and verifying its orientation by polarized-light microscopy (Nikon LV100N)(epi-illumination; analyzer angle set to maximize tidemark contrast; Laser ablation was performed at the Institut Fresnel PHOTONICS facility.

**pSHG and 2PF**
SHG and 2PF images were acquired on a BondXplorer™ commercial scanning microscope (Lightcore Technologies, FR) which contains a SHG acquisition module on which a polarization control was added for polarization resolved SHG (pSHG). pSHG resolves the nano-scale orientational order in the ECM. The SHG signal arises from the non-centrosymmetric arrangement of polypeptide bonds as well as methyl groups along the tropocollagen triple helix, acting as emitters[32–35]. Consequently, the SHG intensity is proportional to the number and degree of co-aligned emitters within the focal volume.
Sections of 5 µm nominal thickness were imaged in epi-geometry using a 25× water-immersion objective (Nikon APO-MP) at an excitation wavelength of 823 nm, from a fs laser oscillator (150 fs width, 80 MHz repetition rate) (Chameleon, Coherent). The laser power at the sample was maintained at 45 mW and kept below the threshold for photodamage throughout. The emission was first reflected by a 705 nm long pass dichroic (FF705-DIO1, Semrock), then spectrally filtered using a short pass filter (FESH0700 Thorlabs), and then split into SHG and 2PF channels using a dichroic mirror (Di02-R488, Semrock). In the detection path, the backfocal plane of the objective is projected on two independent photomultiplier tubes (PMT1001/M, Thorlabs), which detect separately SHG and 2PF signals, further spectrally filtered (SHG 400/40 and 2PF 535/150, Semrock), with bias voltages adapted to the available signal to noise (0.8 V and 1.1 V for SHG and 2PF channels, respectively). Imaging was performed with a pixel dwell time of 20 µs. The field of view comprised 1024 × 1024 pixels at step sizes of 243.75 nm, 325 nm, or 487.5 nm depending on the magnification configuration. For polarimetric measurements, an achromatic half wave plate (AHWP10M-580, Thorlabs) was rotated in a motorized mount (KPRM1E/M, Thorlabs) interfaced with the image acquisition, in order to provide one image per incoming linear polarization state. 18 incoming polarization angles equally spaced between 0° and 180° were recorded per field of view. The pSHG setup was calibrated with a KTP crystal of known orientation, ensuring the absence of polarization distorsions and the setting of the horizontal direction angle. For pSHG/2PF measurements, sample sizes of animals were $N = 5$ (control), $N = 5$ (unloaded), and $N = 3$ (reloaded). pSHG and 2PF microscopy was performed at the Institut Fresnel PHOTONICS facility.

**TexTOM and XRD-CT**

Texture tomography experiments and 3D X-ray diffraction tomography were conducted at the nano-branch (EH3) of beamline ID13 at the European Synchrotron Radiation Facility (ESRF-EBS). Monochromatic X-rays (15 keV) were selected using a Si(111) channel-cut monochromator and pre-focused by beryllium compound refractive lenses onto a set of multilayer Laue lenses[60], yielding a final beam size of 300 × 300 nm² (FWHM) at the sample position.

Samples were scanned using a piezoelectric stage (P-561.3CD, PIMars, Physik Instrumente; 100 µm travel range) in continuous scanning mode and 2 ms exposure time. An in-house goniometer based on SmarAct actuators[61,62] enabled rotation about the z-axis ($\omega$) and tilting about the y-axis ($\kappa$) and was mounted directly on the scanning stage. Diffraction patterns were recorded in transmission geometry using an Eiger 4M detector positioned at a calibrated sample–detector distance of 238.6 mm. Calibration of detector distance to sample as well as detector geometry (beam center, detector tilt) was performed using an $Al_2O_3$ standard (NIST SRM 674a). With a 200 µm diameter lead beam-stop, the accessible q-range was 0.35–26.5 $nm^{-1}$.

The total absorbed dose $d$ was estimated according to

$$d = \frac{\mu N_0 \varepsilon}{\rho},$$

where $\mu = 10.2\ \text{cm}^{-1}$ is the linear absorption coefficient determined from transmission measurements[63], $N_0$ the total incident flux per unit area, $\varepsilon = 2.4 \times 10^{-15}$ J the photon energy at 15 keV, and $\rho = 1.5\ \text{g cm}^{-3}$ the mass density obtained from holotomographic reconstruction. Dose values per specimen are given in Table 1.

The number of projections per tomogram was chosen according to the Nyquist–Shannon sampling criterion, based on the pillar diameter and scan step size. The zero-tilt angle tomogram covered 180°. For tilt angles ≠ 0°, a 360°rotation was acquired, however the sampling of non-zero tilt angles were factorized by cos(θ) to ensure approximately uniform sampling of orientation space while avoiding oversampling at high tilt angles. The complete 3D orientation space was reconstructed from series of 10 tilt-tomograms at $\kappa$= 0°, 5°, 10° … 45°

Because pillar diameters exceeded the 100 µm piezo travel range, each projection was acquired in four overlapping patches (overlaps of 5 µm and 7 µm for step sizes of 0.5 µm and 0.75 µm, respectively); adjacent patches were stitched before reconstruction. Radiation damage was monitored by recording a reference projection at $\omega = 0$, $\kappa = 0$ after each tilt-tomogram series and comparing it to the initial reference; the dose-dependent signal evolution is shown in SI Fig. 9.

| 3D Specimen | Number of projections | Number of 0-tilt projections | Voxelsize [µm] | Incident photon Flux [ph/s] | Dose (TexTOM) [Gy] | mean diameter [µm] | Height [µm] |
|---|---|---|---|---|---|---|---|
| CTL | 295 | 100 | 0.5 | ~$1 \times 10^{12}$ | $5.08 \times 10^{9}$ | 110 | 60 |
| UL | 310 | 100 | 0.5 | ~$5 \times 10^{11}$ | $2.68 \times 10^{9}$ | 80 | 67 |
| RL_1 | 263 | 75 | 0.75 | ~$5 \times 10^{11}$ | $2.28 \times 10^{9}$ | 117 | 87 |
| RL_2 | 261 | 63 | 0.75 | ~$5 \times 10^{11}$ | $2.26 \times 10^{9}$ | 72 | 80 |

*Table 1 Experimental parameters per 3D specimen*

**X-ray microdiffraction**

µXRD measurements were performed at the microfocus end-station of ID13 at the European Synchrotron Radiation Facility (ESRF-EBS) in 16 bunch mode. The incident beam was energy-selected by a channel-cut Si(111) monochromator and subsequently focused with beryllium compound refractive lenses to a spot size of 2 × 3 µm² (horizontal x vertical, FWHM) at the sample position, with the photon energy set to 17 keV. The photon flux at the sample was ~$8.1 \times 10^{11}$ photons $s^{-1}$. An ionization chamber positioned upstream of the specimen was used to monitor incident beam intensity. A 200 µm beamstop was used to block the direct beam. Scanning was performed in continuous mode with 50/100 ms per exposure for SAXS/WAXS , at 2 µm step-size.

SAXS patterns were recorded in transmission geometry using an Eiger X 4M detector positioned at a calibrated (AgBh calibrant) sample–detector distance of 965.2 mm, yielding a q-range of 0.167–5.0 $nm^{-1}$. For WAXS measurements, the detector was positioned at a calibrated ($Al_2O_3$ standard (NIST SRM 674a)) sample–detector distance of 128.5 mm. yielding a q-range of 0.224–43 $nm^{-1}$. Diffraction images were corrected for solid angle, polarization and dead pixels and azimuthally integrated using a bin width of 0.05 $nm^{-1}$ to obtain one-dimensional diffractograms using pyFAI[64]. Sample sizes of specimen were $N$ = 5 (control), $N$ = 6 (unloaded), and $N$ = 7 (reloaded), corresponding to 4, 4, and 5 animals per group.

**Holotomography** data was acquired at nano-imaging beamline ID16A (ESRF). Here, a set of KB-mirrors focus the beam with low divergence. The photon energy was set to 17.1 keV with a measured photon flux of $4 \times 10^{11}$ ph/s. 2000 images were acquired (at four distances to account for zeros in the contrast transfer function) in cone-beam configuration with a propagation depth of $z_{12} = 1.2\ m$. The effective pixel size via geometric magnification was 30 nm. The sample was measured in vacuum at room temperature. Projection images were recorded with a lens coupled XIMEA sCMOS detector (6144x6144 pixels at 10 µm pixelsize, rebinned to 2048x2048 pixels) at 0.25s exposure time along with 50 empty beam background and 20 dark fields for data correction purposes. Quantitative phase-retrieval was performed with a CTF-based reconstruction algorithm, followed by a tomographic reconstruction using the *Nabu* software package (ESRF). To account for zeros in the contrast transfer function, tomographic scans with 2000 projections were performed at four distances[65]. Ring artifact removal was performed based on a combined wavelet-fourier filtering[66]. The mass density was retrieved from the refractive index by assuming a constant electron-to-mass ratio ($Z/A \approx 0.5$) for soft tissue[67]. The total dose that was subjected to one sample for a tomographic reconstruction was $D \sim 5 \cdot 10^7$ Gy.

**Data analysis**

**pSHG**:

Collagen orientation and order parameters were extracted from polarimetric SHG stacks using PyPOLAR(v2023.2.6.2; RRID:SCR_024681; https://github.com/cchandre/Polarimetry.git ), which implements a direct analysis of the SHG polarization response at each pixel. The dependence of the SHG intensity as a function of the variable incoming polarization angle $\alpha$ is analyzed by direct projection on circular basis functions $(\cos n\alpha, \sin n\alpha)$ with $n = (0,2,4)$ [34,68]. From this analysis, the deduced circular decomposition coefficients $(a_{n=0,2,4}, b_{n=2,4})$ can be extracted as the weights of the $(\cos n\alpha, \sin n\alpha)$ functions respectively, defining the intensity decomposition amplitudes $I_{n=2,4} = \sqrt{a_n^2 + b_n^2}/a_0$. These coefficients can be expressed as functions of the order parameter accessible by pSHG, which is the ratio between the third and first symmetry orders of the distribution function of the dipoles generating SHG in the sample[34,68]. In collagen, radiating dipoles at the source of SHG signals lie along the peptide bonds that constitute the collagen filaments helix shapes, within the collagen macromolecular organized structure. The order parameter, which quantifies the angular extent explored by these dipoles, is therefore a signature of the organization of collagen at the molecular level, averaged over the focal volume of the SHG microscope (typically hundreds of nanometers dimension). This average therefore embeds the molecular order within collagen fibrils, and order formed by these collagen fibrils into larger fiber structures. The resulting order parameter called $S_{SHG}$ varies between $1$ and $-3$ for an angular aperture of the dipoles between 0° (dipoles all parallel) to 180° (dipole following an isotropic distribution) [34,68], i.e. $S_{SHG}$ decreases with a decreasing molecular order. Typically, $S_{SHG} = -0.4$ in single collagen I fibers [34,68]. It can be shown that for a large range of intermediate aperture angles, $S_{SHG}$ can be extracted from the circular amplitude coefficients $I_{n=2,4}$ by a simplified interpolated function $S_{SHG} = -\frac{1}{2}.((I_2 - I_4)/(I_2 + I_4) + 1.3)$, which makes the analysis computationally faster.

Background was identified by histogram thresholding and subtracted to the polarization responses before analysis. Spatial binning was applied iteratively until the model–data residual fell within the Poisson shot-noise level, requiring a binning factor of 3 or 5 depending on the section.

Segmentation used the combined pSHG total-intensity and 2PF images. The tidemark was manually delineated as a one-pixel-wide reference line; background pixels and distal regions outside the section plane were excluded (SI Fig. 10).

For each image, pixel values were binned by Euclidean distance from the tidemark in 3 µm bins (reference pixel size 0.456 µm to unify step sizes across sections). Specimens imaged as overlapping patches were combined into a single specimen-level profile by weighted averaging on pixel count per bin; the animal was treated as the statistical unit throughout. Group profiles were obtained by averaging specimen-level profiles; the standard error of the mean (s.e.m.) was computed per distance bin.
The 2PF signal was normalized per image by the mean of the background subtracted signal in the UFC, in the first 50 µm away from the tidemark. The SHG signal was normalized per image by the mean signal in the UFC, in the first 50 µm away from the tidemark.

**X-ray microdiffraction processing**Mineral-containing pixels were identified by multi-Otsu thresholding of the azimuthally integrated diffraction intensity; non-mineral pixels were excluded from Rietveld refinement and SAXS fitting. The tidemark was manually segmented (See SI Fig. 11). Per specimen, one distance-resolved profile was extracted by binning pixel values by Euclidean distance from the tidemark; group profiles and s.e.m. were computed as for the optical data.

**XRD-CT data processing**

Projections were aligned using an optical-flow based registration approach[69] (Mumott version 2.2), applied to the mean scattering signal in the SAXS region $q$ = 0.8 - 1.14 nm$^{-1}$, a region where the PMMA contribution is weak compared to the bone mineral scattering, enabling sub-pixel alignment precision. XRD-CT voxel diffractograms were reconstructed per $q$-value (2000 points from 0.35 – 26.5 nm$^{-1}$), by normalizing first by the ion chamber, then scaling to [0, 1] and then applying SIRT[69] reconstructions (100 iterations). Identical processing parameters were used for all groups.

**SAXS data fitting**

The $T$-parameter ($T$) was estimated from SAXS data assuming a two-phase system predominantly platelet-shaped mineral particles [38,70]. The $T$-parameter was estimated from the Porod constant $P$ and invariant $J$ using, as previously described [61,71]:

$$T=\frac{4J}{\pi P}=\frac{4}{\pi P}\int_0^\infty q^2 I(q)\,dq,$$

Where $I(q)$ denotes the position-resolved scattering intensity. The Porod constant was obtained by linear extrapolation of the Porod plot, $I(q^4)$, from the $q$-range $1.25 - 2.1\ \mathrm{nm}^{-1}$ to $q = 0$. The invariant $J$ was determined from the Kratky plot, i.e. $I(q^2)$, by integrating $q = 0.35 - 2.0\ \mathrm{nm}^{-1}$ and extrapolating towards $q = 0$ and $q \to \infty$ using a $q^{-4}$ dependence.
For the µ-XRD dataset, the same procedure was applied over a $q$-range 0.35-2.0 $\mathrm{nm}^{-1}$, using $q > 1.16\ \mathrm{nm}^{-1}$ for Porod extraction. Because T depends jointly on mineral particle size and degree of mineralization, it is used throughout as a nanostructural proxy rather than a direct particle-size measurement.

**Rietveld refinement:**

Sequential Rietveld refinement of each voxel/pixel diffractogram was performed using MultiRef[72], a Matlab-based interface to GSAS[73]. An initial structural model was derived from a representative voxel and then applied to the entire dataset. The bioapatite model was based on geological hydroxyapatite (Crystallography Open Database 9011096). No instrumental broadening was included, as crystallite sizes in bone and mineralized fibrocartilage are sufficiently small that peak widths are dominated by size broadening.

Refinement proceeded in four sequential steps: (1) sixth-order Chebyshev background and scale factor; (2) preferred orientation using spherical harmonics to order 6; (3) lattice parameters; (4) peak profile parameters, with all parameters refined jointly in the final cycle. Peak shapes were modelled by a pseudo-Voigt function incorporating isotropic and anisotropic size broadening; the *c*-axis was treated as the principal anisotropic direction and the crystallite modelled as a rotationally symmetric prolate ellipsoid parameterized by length and diameter. Voxels showing no discernible diffraction signal were excluded.

**TexTOM**:
Diffraction patterns were rebinned into 100 radial ($q$) and 120 azimuthal bins over the q = 8.06 - 26.41 $nm^{-1}$. The (002) reflection $q$ = 17.66 - 18.62 $nm^{-1}$ and a multiplet comprising the (21-31), (11-22), (30-30) and (20-22) reflections $q$ = 21.50 - 24.39 $nm^{-1}$were selected from a representative azimuthally integrated diffraction profile. Background subtraction was performed using a linear baseline fitted to peak-free regions adjacent to the chosen $q$-regions.
Texture tomograms were reconstructed using TexTOM[26] The three-dimensional crystallite orientation distribution function (ODF) was parameterized by a hyperspherical harmonic expansion up to order 20 and optimized using projected gradient descent. Nematic order parameters and directors were obtained from the largest eigenvalue and corresponding eigenvector of the second-rank orientation tensor[74] $\mathbf{Q}$

$$\mathbf{Q} = \int \mathrm{ODF}\,(g') \left(\frac{3}{2}\hat{\mathbf{c}}(g') \otimes \hat{\mathbf{c}}(g') - \frac{1}{2}\mathbf{I}\right) dg'$$

where $g'$ denotes crystal orientation, $\hat{\mathbf{c}}$ the crystallographic $c$-axis unit vector, $\mathbf{I}$ the identity tensor and $\otimes$ denotes the tensor product, whereas the director describes the average crystallographic orientation. Local misalignment was calculated as the mean angular deviation between the director of a voxel and those of its neighboring voxels N within a $3 \times 3 \times 3$ window.

$$\Delta\theta_i = \frac{1}{N}\sum_{j=1}^{N} \arccos\,\left(\left|\mathbf{n}_i \cdot \mathbf{n}_j\right|\right),$$

with $\mathbf{n}_j$ the jth nematic director.

**3D Structure tensor analysis**

The reconstructed holotomographic mass density volumes were analyzed using a local structure tensor to quantify the directional organization of mineral density variations. The structure tensor, computed from local image gradients, is sensitive to oriented features such as planar interfaces and elongated structures. Its eigenvalues and eigenvectors describe the magnitude and principal directions of local anisotropy, with the smallest eigenvalue corresponding to the direction of least intensity variation. For quantitative comparison, local anisotropy was summarized by the fractional anisotropy (FA), which measures the degree to which density variations are preferentially oriented. High FA values indicate strongly anisotropic organization, whereas low FA values correspond to more isotropic structures[75,76].

Where $I(\vec{x})$ is the mass density at position $\vec{x} = (x, y, z)$, the image was first smoothened by convolution with an isotropic Gaussian kernel $G_\sigma$. This yielded the smoothened image $I_\sigma = G_\sigma * I$, where $*$ denotes a convolution. The partial derivatives $\partial_i I_\sigma$ were estimated numerically over a one-voxel neighborhood. From a cubic window $W$ of side length $w$, mean the local structure tensor is defined as:

$$J_{ij} = \langle (\partial_i I_\sigma)(\partial_j I_\sigma) \rangle,$$

with $\langle \cdot \rangle$ the mean over all voxels contained in $W$. This is thus the mean of the gradient's outer product with itself inside of Window $W$. This symmetric tensor $J$ was diagonalised to obtain the eigenvalues:

$$\lambda_1 \leq \lambda_2 \leq \lambda_3.$$

These eigenvalues quantify the mass density variation along the principal directions of the local microstructure ellipsoid.
From this the directional organization was quantified using the fractional anisotropy (FA):

$$FA = \sqrt{\frac{3}{2} \frac{\sum_{i=1}^{3} (\lambda_i - \bar{\lambda})^2}{\sum_{i=1}^{3} \lambda_i^2}},$$

with $\bar{\lambda} = \frac{1}{3} \sum_{i=1}^{3} \lambda_i$.
Here, high FA values correspond to locally aligned mineral interfaces and tessellated mineral structures, whereas low FA values indicate isotropic mineral organization. The FA was therefore used as a quantification of local directional organization of the mineral interfaces, yielding a quantitative measure of texture in the mineralized tissue.
The Gaussian smoothening parameter ($\sigma$) and structure tensor window size ($w$) were optimized independently. The smoothing parameter was evaluated by comparing the stability of the fractional anisotropy (FA) histograms for $\sigma$ values ranging from 0.8 to 1.8 voxels. A value of $\sigma$=1 voxel was selected as it produced stable FA histograms while preserving the underlying microstructural features. The window size $w$ was evaluated by comparing the histogram of FA over the samples MFc region for different window sizes (30, 40, …, 120 pixels). The histogram stabilized at window size of 80 pixels, which corresponds roughly to 2-3 times the characteristic size of a mineral tessel. The window operation was applied at a stride of 10 voxels in each dimension.

**Post-processing and Segmentation of 3D data**

The tidemark was manually delineated across the mineralization front in both the TexTOM/XRD-CT and holotomography datasets. A one-voxel-thick tidemark shell was constructed by subtracting the original segmentation mask from a morphologically dilated version and restricting the result to the manually defined tidemark region. Euclidean distance maps from this shell provided the reference coordinate for all distance-resolved analyses.

Holotomography volumes were segmented in Dragonfly 3D World (v2024.1, Object Research Systems, Montreal, QC, Canada) using local histogram thresholding (local Otsu with adaptive Gaussian weighting) to separate mineralized from soft tissue (SI Fig. 12).

TexTOM nematic director fields were visualized as streamplots in Paraview(v5.13.2)[77], seeding 5000 streamlines in a 100 µm-radius sphere centered on the tidemark and integrating with a fourth-order Runge–Kutta scheme until convergence.

**Data availability**

The datasets for the 2D µXRD are available under DOI https://doi.org/10.15151/ESRF-DC-2454674292, the datasets for the 3D TexTOM/XRD-CT are available here https://doi.org/10.15151/ESRF-DC-2459912452 and the 3D Holotomography data is available here https://doi.org/10.15151/esrf-dc-2484024140

The TexTOM code package used for the analysis of the TexTOM/XRD-CT data is available under https://gitlab.fresnel.fr/textom/textom

**Author contributions**

TG, and MP conceived the project, TG, SR, MS designed the experiment with input from CC and MP. MS, NW, HB, CC, TF, CG, AK, KI, TB, AC and SR prepared the samples, MS, MF, ISB, ME, RR, MB, SB, SR and TG performed experiments, MS, MF, ISB ,CG, ME, MB, SR and TG performed the synchrotron experiments. MS, MF, SB, HB and TG developed the data analysis software, MS, MF, CC, SB, HB, MP, SR and TG analyzed the data. TG supervised the project, TG and MP acquired the funding. TG, MS wrote the initial draft of the manuscript with input from SR, CC and HB and revised it with contributions from all authors

**Competing Interests**

The authors declare no competing interests.

**Acknowledgements**

We acknowledge the ESRF for supplying beam time for proposals LS-3153, LS-3299, CH-7039 and ID16A for providing in-house experimental time and the Partnership for Soft Condensed Matter (PSCM) for support during the preparation of the experiment. We thank Pierre Paleo for help with ring artifact removal The pSHG/2PF microscope facility has received funding from the France 2030 investment plan managed by the French National Research Agency (ANR), through the IDEC Equipex+ grant (France 2030 investment plan ANR-21-ESRE-0002).

This work is funded by the European Union, European Research Council Horizon Europe (grant No.101041871). Views and opinions expressed are those of the author(s) only and do not necessarily reflect those of the European Union or the European Research Council. Neither the European Union nor the granting authority can be held responsible for them.
AC and SR acknowledge funding from the French Centre National d'Etudes Spatiales (CNES, 4800000797) for funding the animal experiments.
We acknowledge help from the Institut Fresnel PHOTONICS facility during the laser ablation preparation of the samples and the polarized SHG and 2PF measurements.
We thank the staff from the METAMUS DMEM facility (https://ror.org/00xffm983) for help with help with the animal experiments. This facility belongs to the Montpellier Animal Facilities Network (RAM, Biocampus).
Support from the Danish Agency for Science, Technology, and Innovation (DanScatt) is gratefully acknowledged. In-house µCT was performed using equipment acquired through the Novo Nordisk Foundation research infrastructure AXIA (grant NNF19OC0055801). We acknowledge support from the ESS lighthouses on hard materials in 3D, SOLID, funded by the Danish Agency for Science and Higher Education, grant number 8144-00002B, and the ESS lighthouses for Multiscale Structural Biology with Neutrons and Data Science in Life Sciences, NeuData4Life, funded by the Danish Agency for Science and Higher Education, grant number 5312-00001B.

**Supplementary information for**

**The Achilles tendon enthesis rebuilds its mineralization front on reloading but retains a nanoscale imprint of unloading.**

M.L. Stammer[1], C. Camy[2,3], M. Frewein[1], I. Silva Barreto[1], C. Genovesio[4], M. Eckermann[5], A. Karimbana[1], K. Iliopoulos[1], R. Ranjan[1], N. Wittig[6], T. Fovet[7], T. Brioche[7], A. Chopard[7], M. Burghammer[5], S. Brasselet[1], H. Birkedal[6], M. Pithioux[2,3,8], S. Roffino[2]‡, T.A. Grünewald‡[1,*]

[1] Aix-Marseille Université, CNRS, Centrale Med, Institut Fresnel, Marseille, France

[2] Aix-Marseille Université, CNRS, ISM, Institut des Sciences du Movement, Marseille, France

[3] Aix-Marseille Université, APHM, CNRS, ISM, Mecabio Facility, Anatomy Laboratory, Timone, Marseille, France

[4] Aix-Marseille Université, Faculté de Pharmacie, Marseille, France

[5] European Synchrotron Radiation Facility, Grenoble, France

[6] Aarhus University, Department of Chemistry, Aarhus, Denmark

[7] PhyMedExp, Univ Montpellier, Inserm, CNRS, Montpellier, France

[8] Aix-Marseille Université, APHM, CNRS, ISM, Sainte-Marguerite Hospital, Institute for Locomotion, Marseille, France

‡ equally contributing
*Correspondence: tilman.grunewald@fresnel.fr

**Content**

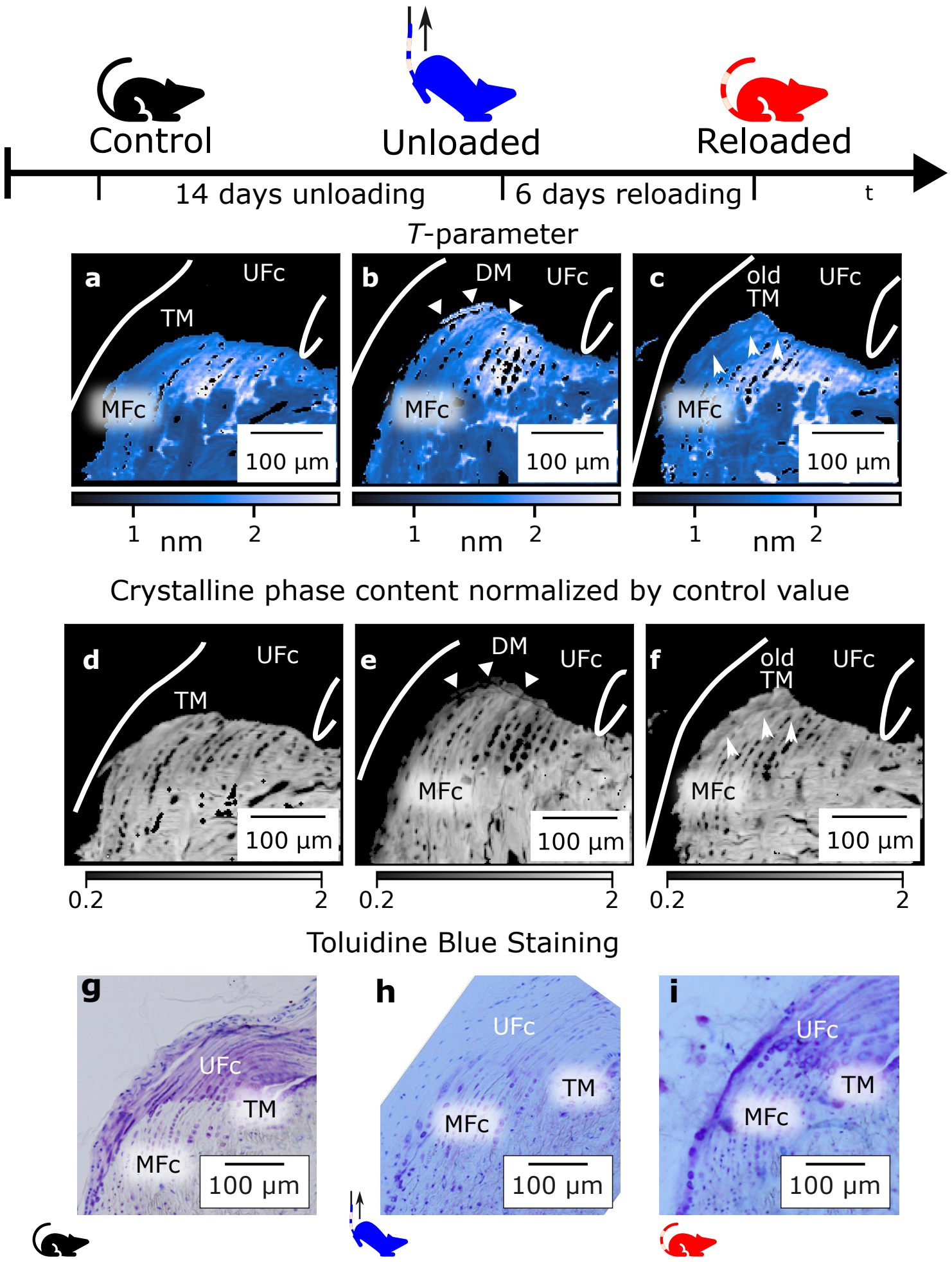

*SI Figure 1* ***Effects of unloading on nano-scale, ECM proteoglycan content and crystalline phase content**. **a-c** nano-scale organization changes at the tidemark, quantified by the T parameter via µ-beam Synchrotron X-ray diffraction on sagittal plane sections parameter **d-f** Crystalline phase content quantified by the Scale Factor of the Rietveld refinement profile model, showing a low crystalline tissue beyond the original tidemark after unloading and a discontinuity running through the mineral proper in the reloaded sample, indicating the original tidemark. [DM = diffuse mineralization, triangular arrowheads] [old TM, narrow arrowheads] **g-i** Toluidine blue staining of 2d sagittal plane sections show the strong depletion of proteoglycan content following unloading, which recovers after reloading*

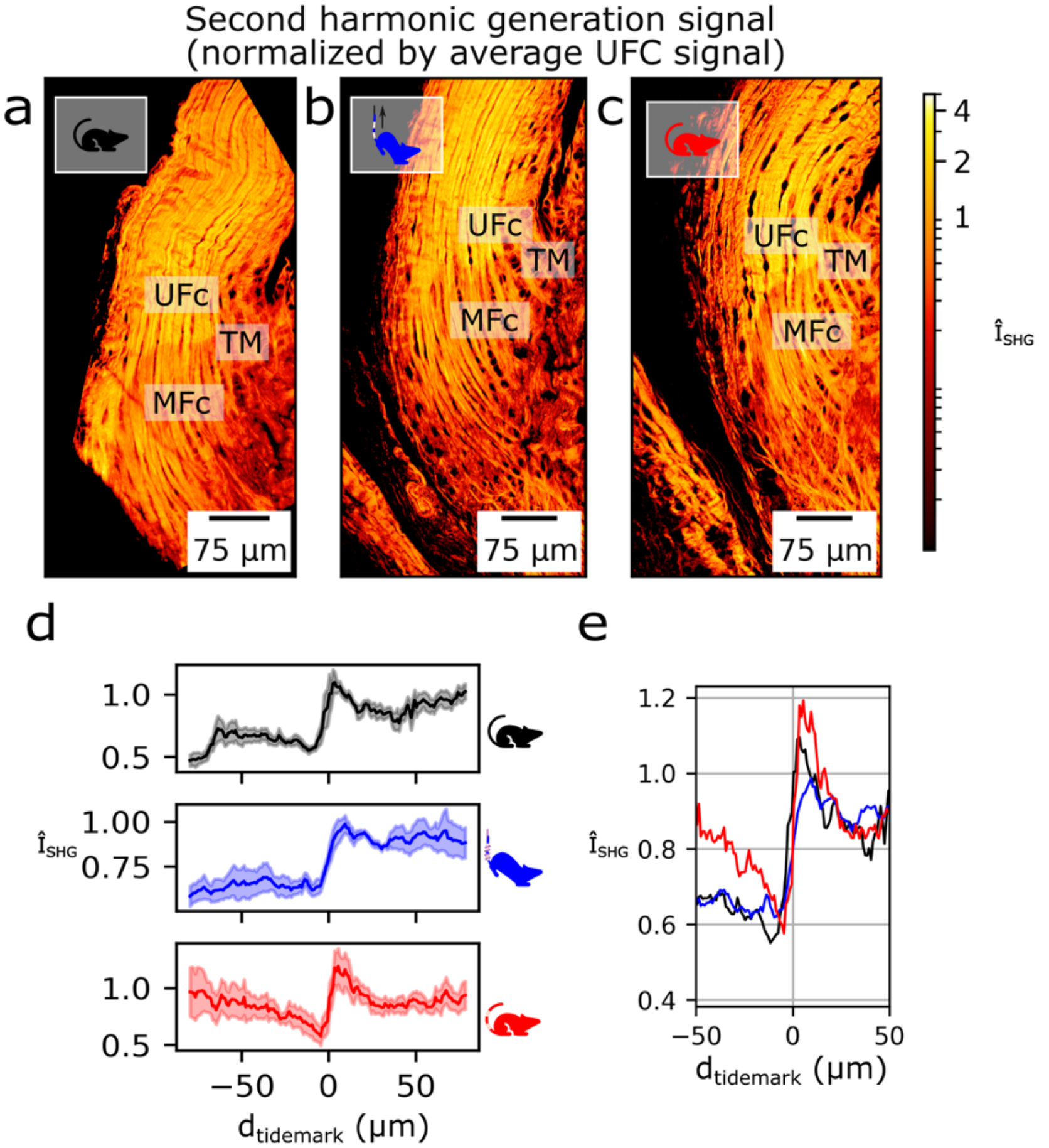

*SI Figure 2* ***Integrated pSGH signal (180° in 10° steps) show the mineralized and unmineralized fibrocartilage exhibit a contrast in SHG-active collagenous tissue. a-c*** *Total integrated SHG signal (normalized by the UFc signal) of example 2D specimen (control, unloaded, reloaded as black, blue and red mice, respectively) plotted logarithmically. UFc and MFc are labeled in the image.* ***d*** *Group averaged lineprofiles of the integrated pSHG signal (normalized by the UFC signal) shown with s.e.m as precision measure. There is a higher variance close to the tidemark in the MFc and a slightly increased signal.* ***e*** *Zoom-in to the 100 µm around the tidemark.*

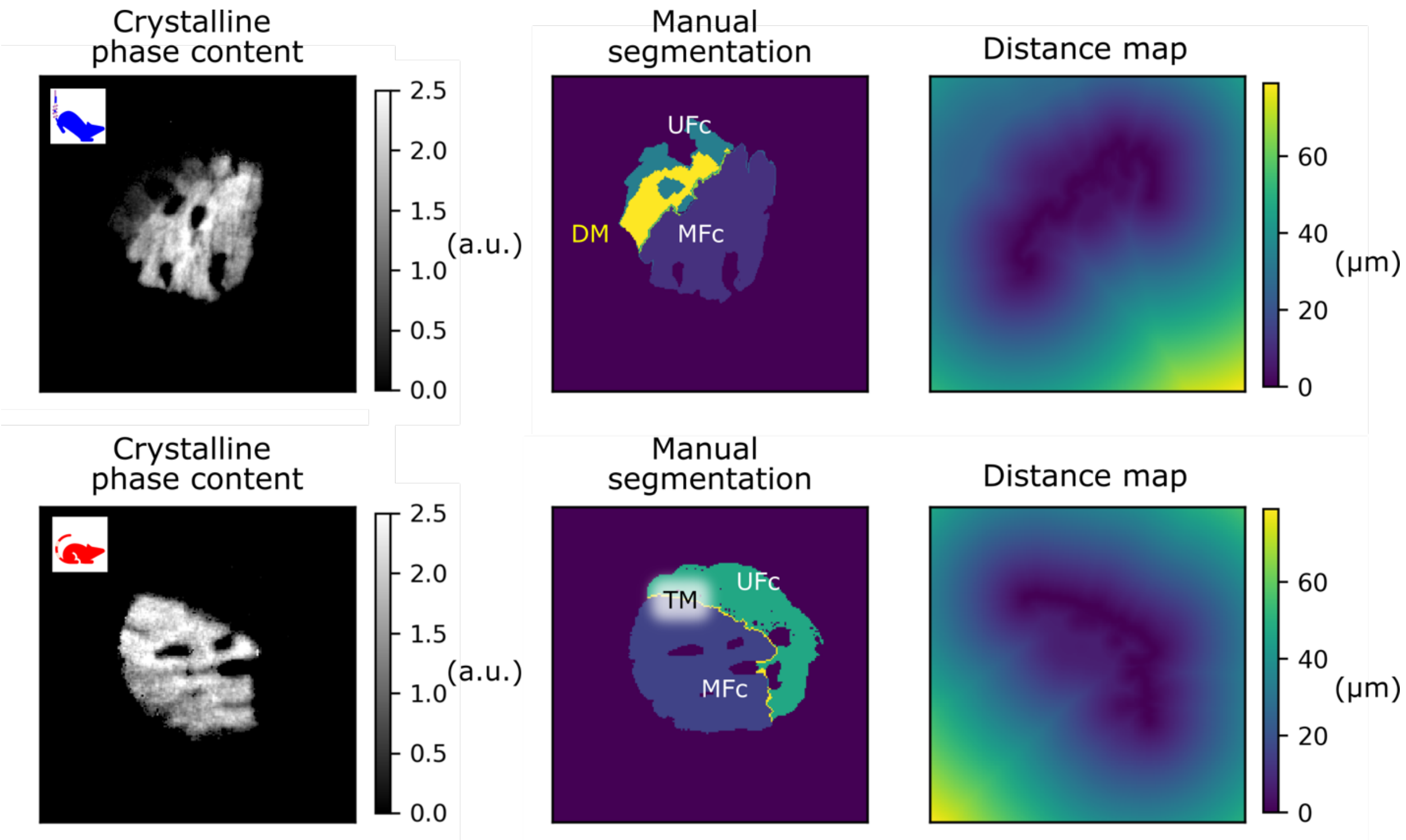

*SI Figure 3* ***Segmentation methodology of 3D nano-diffraction results*** *Rietveld refinement was performed on voxels that exhibited crystalline phase, the scale factor of the profile model then determines original mineralized tissue (MFc) from the diffuse mineralization region (DM). UFc was segmented using the scattering signal from the SIRT reconstructed volume. Tidemark was manually selected from the morphologically determined shell of the (non-diffuse) mineralized tissue. From this 3D surface, the Euclidean map was determined.*

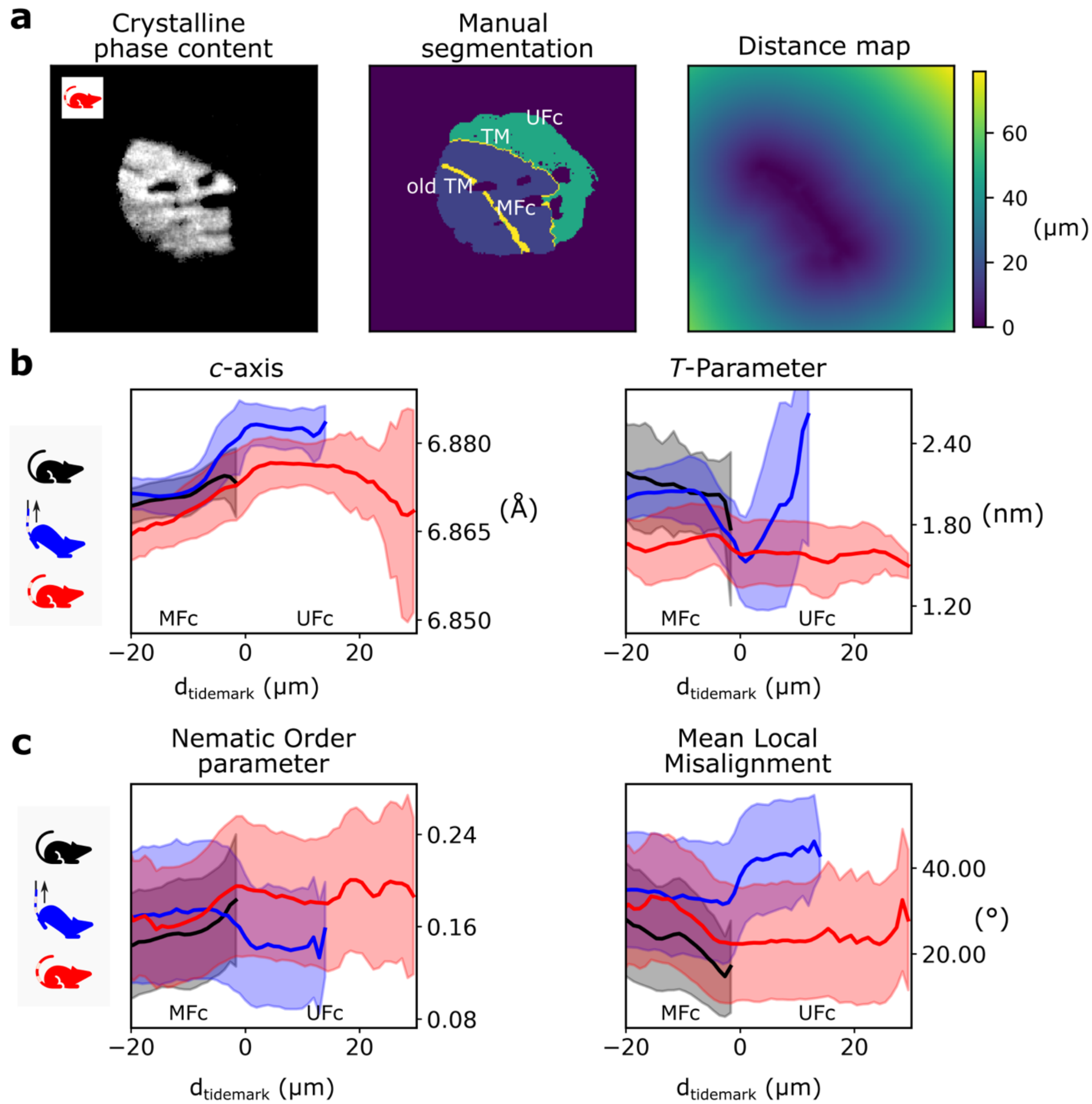

*SI Figure 4* ***Segmentation of old tidemark in RL 3D specimen*** *The old tidemark was visible over multiple retrieved parameters, nano-scale to crystal lattice scale. The old tidemark was segmented manually in 3D and the Euclidean distance map calculated. From this the specimen resolved line plots are retrieved (standard deviation shown). The c-axis shows no recovery after reloading. The T-parameter in lower in the newly formed mineralized fibrocartilage. The crystallite nematic order parameter as well as mean local misalignment show a characteristic order that differs from the unloaded diffuse mineral.*

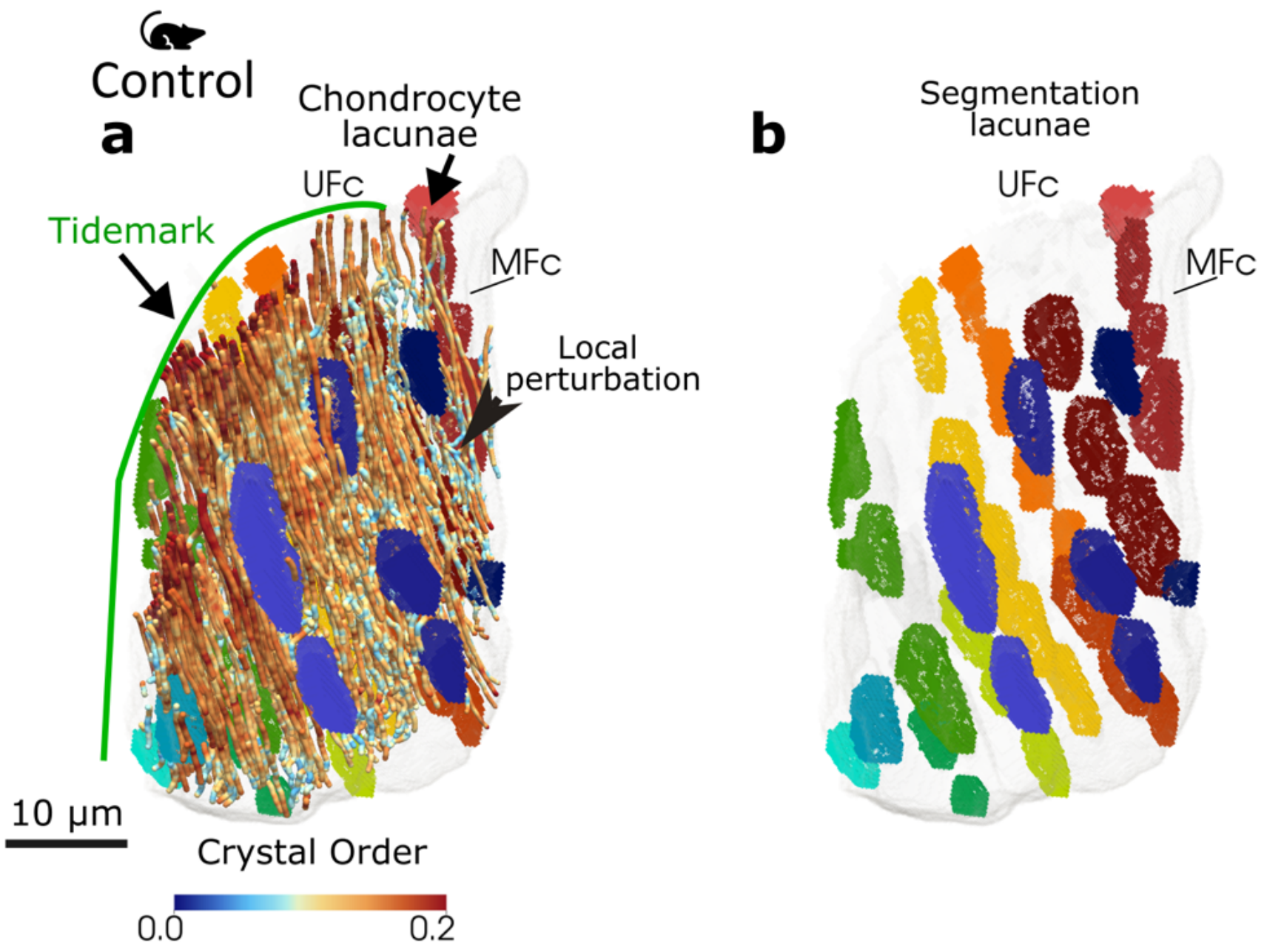

*SI Figure 5* ***Chondrocyte lacunae segmentation, local crystallite perturbation and specimen averaged crystallite properties depending on position relative to chondrocyte lacunae a*** *watershed segmented chondrocyte lacunae, color coded to unify all cells that fall in one row. Arrow points out a local perturbation in the crystallite order due to the lacunae in the mineral proper.* ***b*** *Watershed segmented chondrocyte lacunae, color-coded as above.*

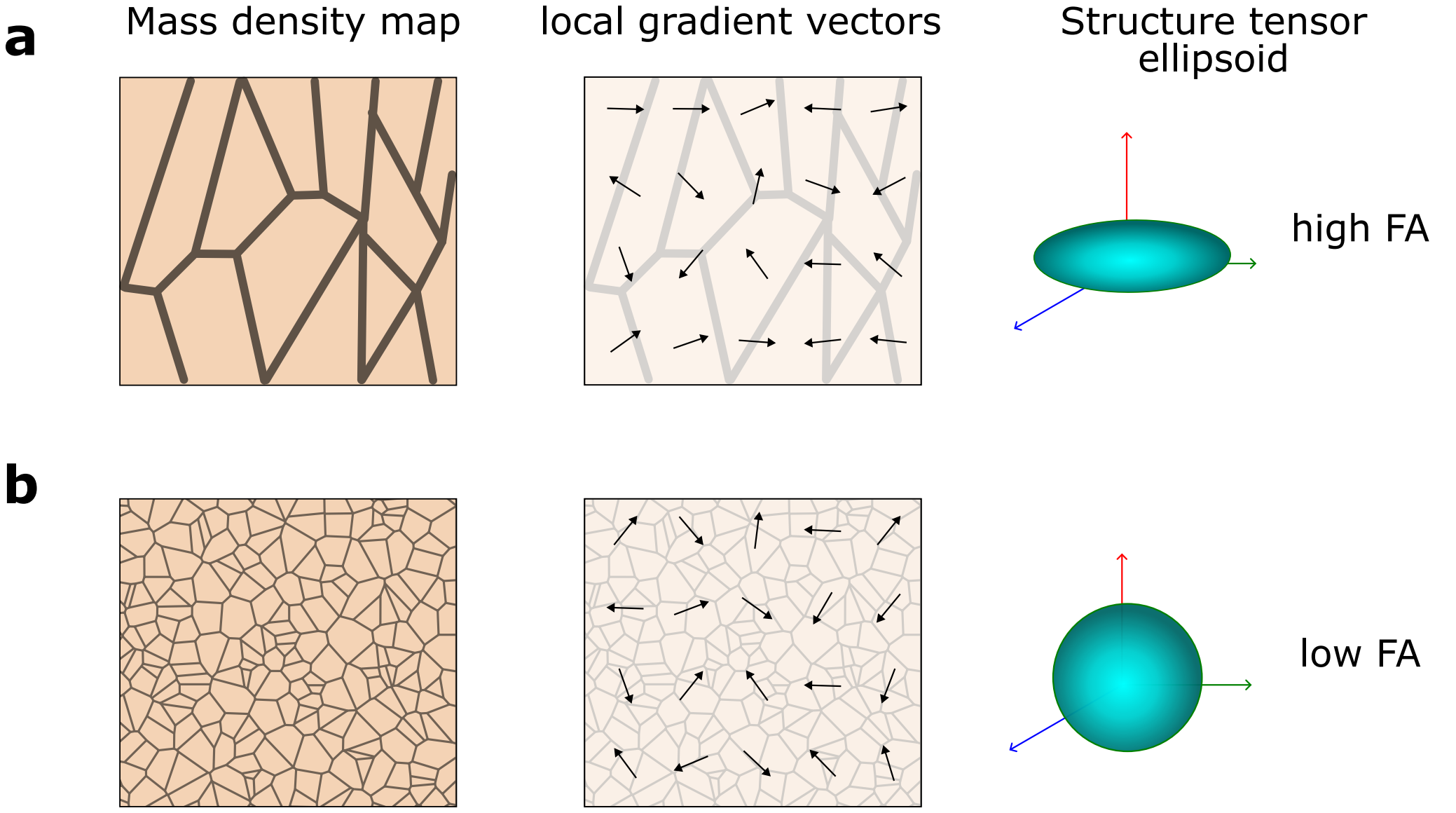

*SI Figure 6* ***Structure tensor visualization, quantifying a change in microstructure mineralized fibrocartilage of the 3D specimen due to unloading a-b*** *Mass density variations due to interfaces result in strong gradients.* ***a*** *For a region that has many aligned iso-surfaces, the averaged structure tensor eigenvalue decomposition yields an ellipsoid, with it's eigenvector corresponding to the lowest eigenvalue pointing in that direction.* ***b*** *Smaller grainsize as well as isotropic texture is reflected in the total FA, yielding a lower value.*

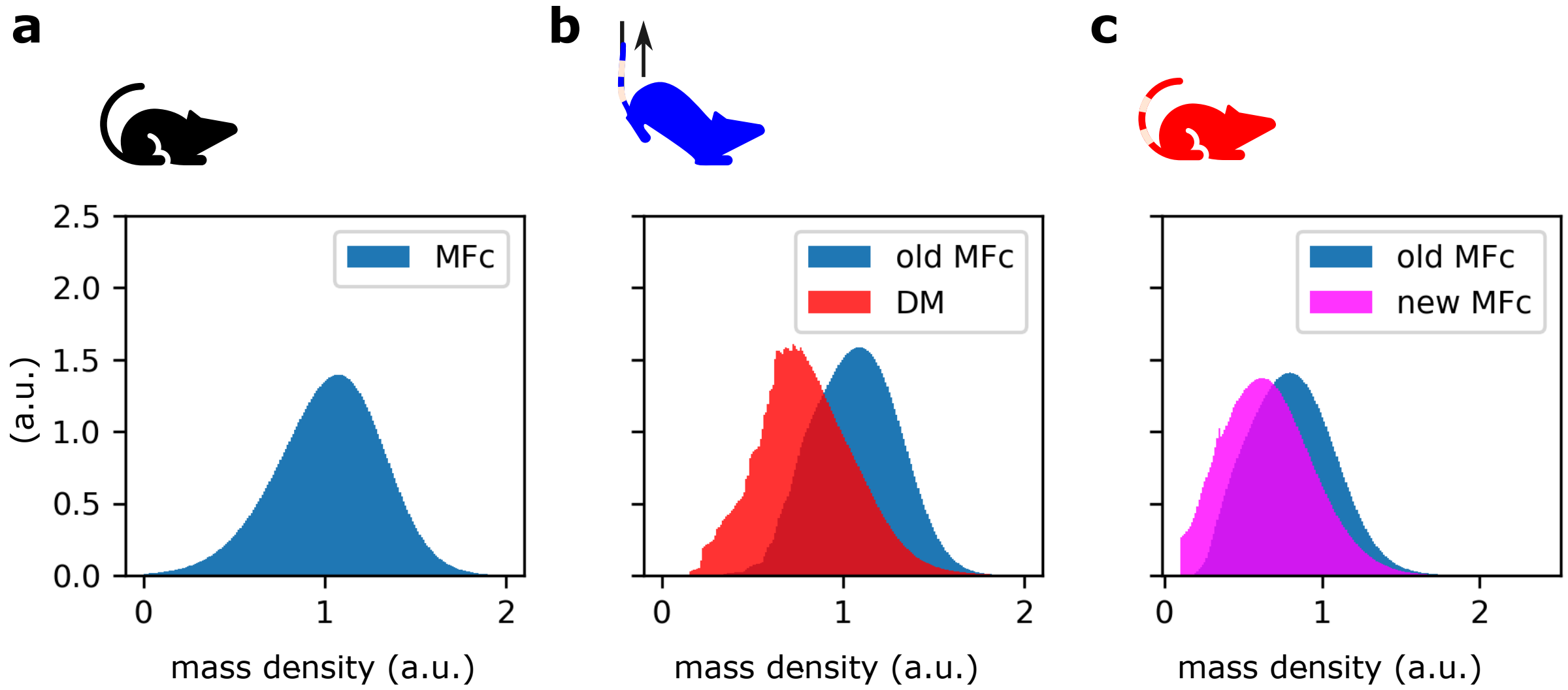

*SI Figure 7* ***Mass density histograms of CTL, UL and RL samples.*** *Relative to the original mineralized fibrocartilage (old MFc), diffuse mineral (DM) in UL and the newly formed mineralized fibrocartilage (new MFc) in RL exhibit distributions that are lower in mass density.*

*SI Table 2* ***Statistical comparison of local fractional anisotropy (FA) and mass density in native mineralized fibrocartilage (MFc), diffuse mineral (DM), and newly mineralized fibrocartilage (new MFc).*** *Values are mean ± s.d. over the structure-tensor window within each specimen. Comparisions are within specimen; D is the Komogorov-Smirnov statistic, reported as a descriptive measure of distributional separation. In RL, new MFc differs from old MFc in both mass density and FA. In UL, the DM differs from the native MFc in mass density whereas the mean FA does not differ but its distribution is broader (D = 0.20)*

| | UL - MFc | UL - DM | Welch's t test | KS *D* | KS *p* |
|---|---|---|---|---|---|
| FA | 0.49 ± 0.09 | 0.50 ± 0.13 | 0.52 | 0.20 | 0.01 |
| Mass density (a. u.) | 1.06 ± 0.23 | 0.82 ± 0.26 | <0.0001 | 0.42 | <0.0001 |
| | | | | | |
| | RL - old MFc | RL - new MFc | Welch's t test | KS *D* | KS *p* |
| FA | 0.53 ± 0.08 | 0.45 ± 0.08 | <0.0001 | 0.46 | <0.0001 |
| Mass density (a. u.) | 0.82 ± 0.26 | 0.67 ± 0.29 | <0.0001 | 0.65 | <0.0001 |
| | | | | | |
| | CTL - MFc | | | | |
| FA | 0.56 ± 0.11 | - | - | - | - |
| Mass density (a. u.) | 1.02 ± 0.29 | - | - | - | - |

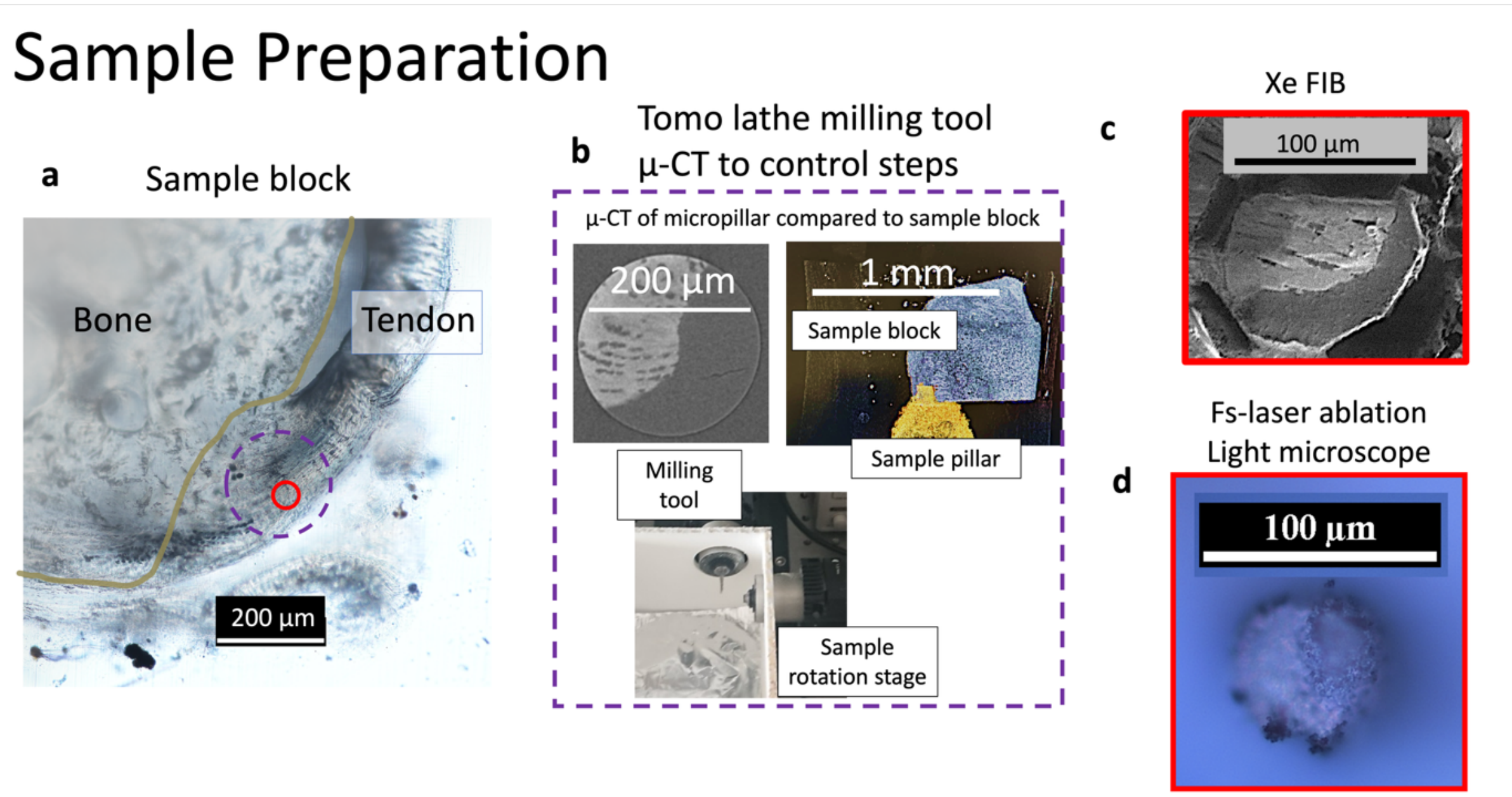

*SI Figure 8* ***3D specimen preparation methodology a*** *Mouse hindlegs, embedded in MMa and sectioned along the sagittal plane until the middle of the foot was reached. From This the ROI was selected.* ***b*** *Milling procedure: micro-pillar was produced via mechanical milling. Milling tool and sample rotation stage were setup in perpendicular geometry. Micro-meter translational stages then allow for cylindrical milling. A procedural µ-CT scan was performed before and after milling to locate the ROI and monitor milling. The original block and the milled sample pillar on top of the MMa sample holder is shown.* ***c*** *Fine-milling with Xe plasma FIB. The micro-pillar of around 200-300 µm diameter was milled down to roughly 100 µm.* ***d*** *Single-shot fs-laser ablation: top view of fine-milled sample (separate sample from d).*

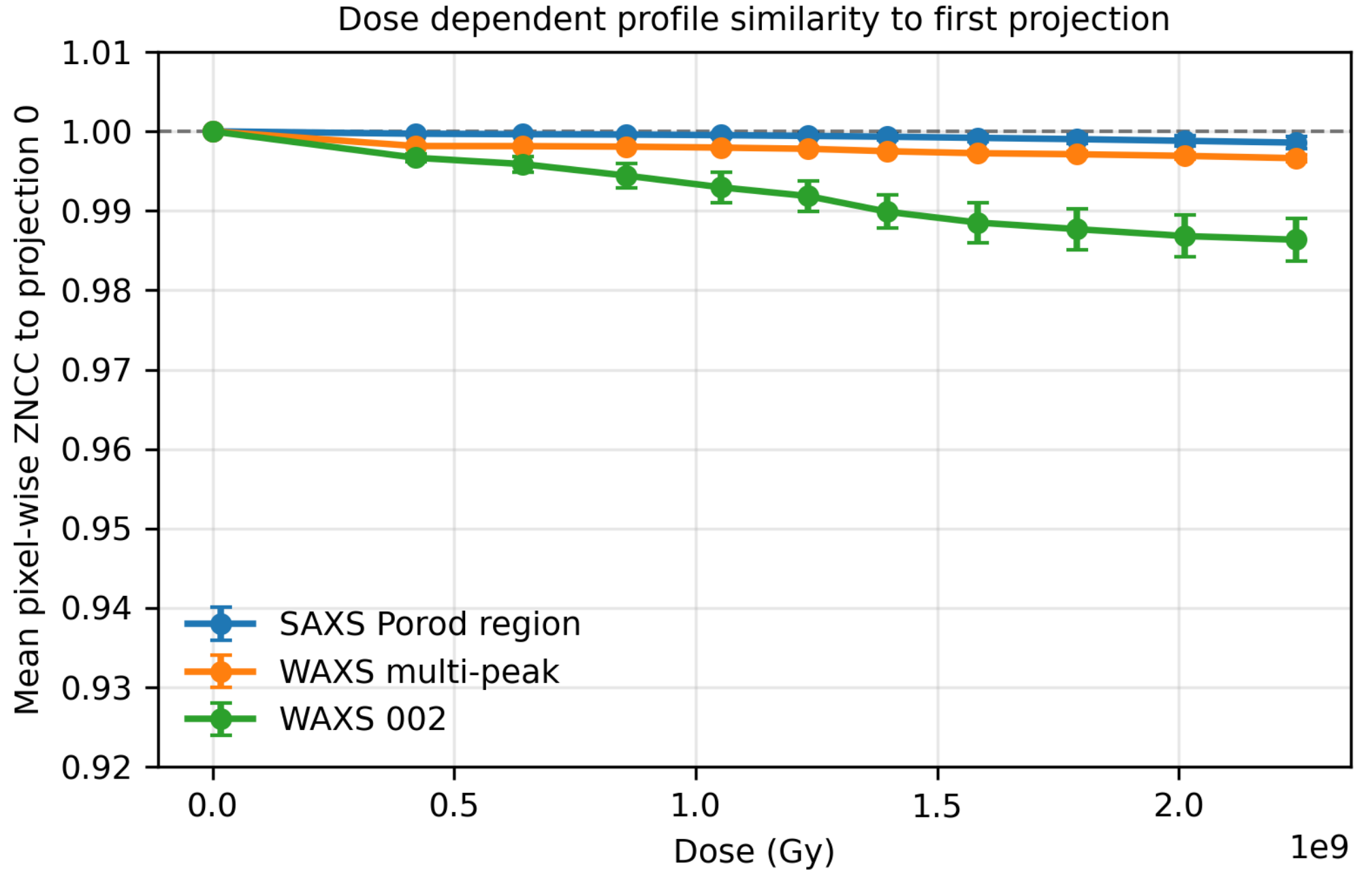

*Figure 9* ***Signal stability monitored by a zero-tilt, zero-angle reference projection after each tilt-tomogram, RL_1 sample****; Mean pixel-wise zero-normalized cross-correlation (ZNCC) was calculated between each projection and projection 0 over selected q ranges corresponding to the Porod region ( 1.0 – 2.0 $nm^{-1}$), The multi-peak region (20.0 – 25 $nm^{-1}$) as well as the (002) peak region (17.7 – 18.8 $nm^{-1}$). The Signal was binned spatially (4x4) before calculating the ZNCC of each binned pixel with the corresponding binned pixel in projection 0. The average was plotted over the corresponding dose with the std as a*

*measure of variance. Empty pixels were excluded. As a preprocessing step, each pixel was normalized by the incoming flux, quantified by an ionization chamber upstream of the sample. For the WAXS signals, the background was subtracted before calculating the correlation.*

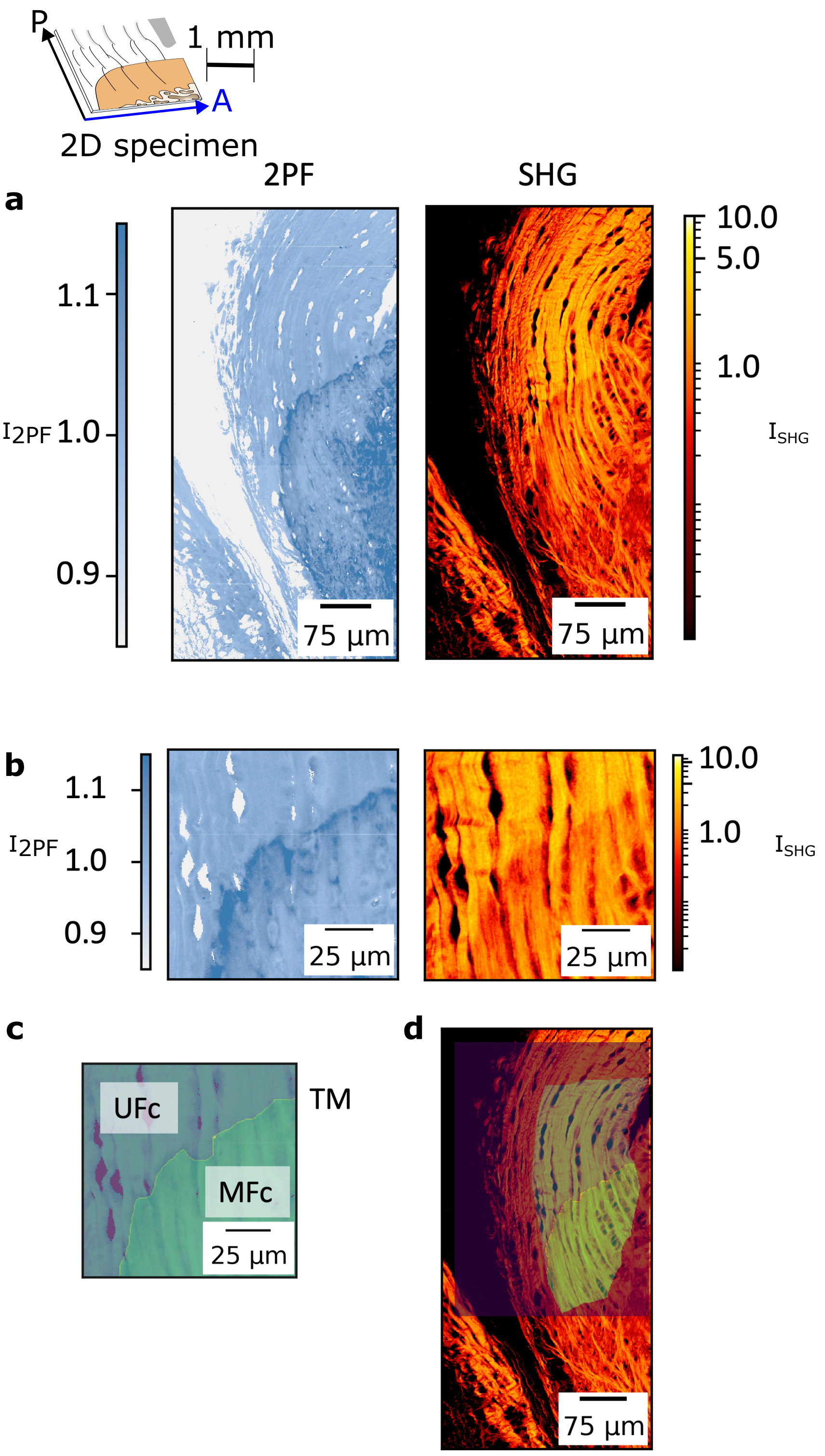

*SI Figure 10* ***Segmentation methodology for non-linear light microscopic sample sections*** *Example total integrated SHG signal (normalized by the UFC signal) from 2D specimen (control, unloaded, reloaded as black, blue and red mice, respectively). UFc and MFc are labeled in the image.* ***d*** *Group averaged lineprofiles of the integrated pSHG signal (normalized by the UFC signal) shown with s.e.m as precision measure. There is a higher variance just distal to the tidemark and a slightly increased signal.* ***e*** *Zoom-in to the 100 µm around the tidemark.*

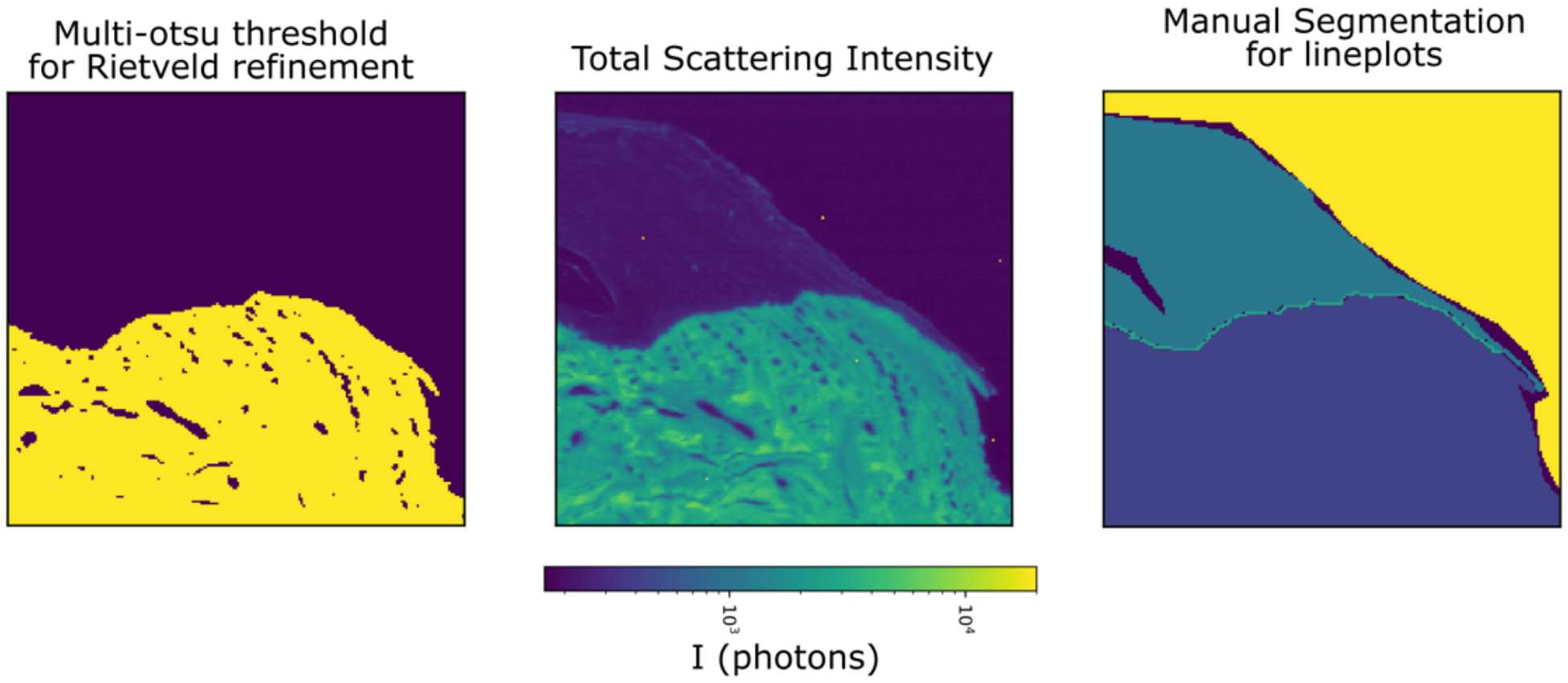

*SI Figure 11* ***Segmentation methodology for the μ-beam Synchrotron X-ray diffraction experiments*** *Starting from the total scattering intensity (q = 0.62 – 50* $nm^{-1}$*) mineralized tissue, soft tissue and background were segmented via a multi-otsu threshold, with which the pixel mask for position-resolved Rietveld refinement was created. The Tidemark and mineralized tissue were segmented manually. SAXS and WAXS scanning geometries were aligned such as to use the same mask. These masks were then used to create the Euclidean distance maps.*

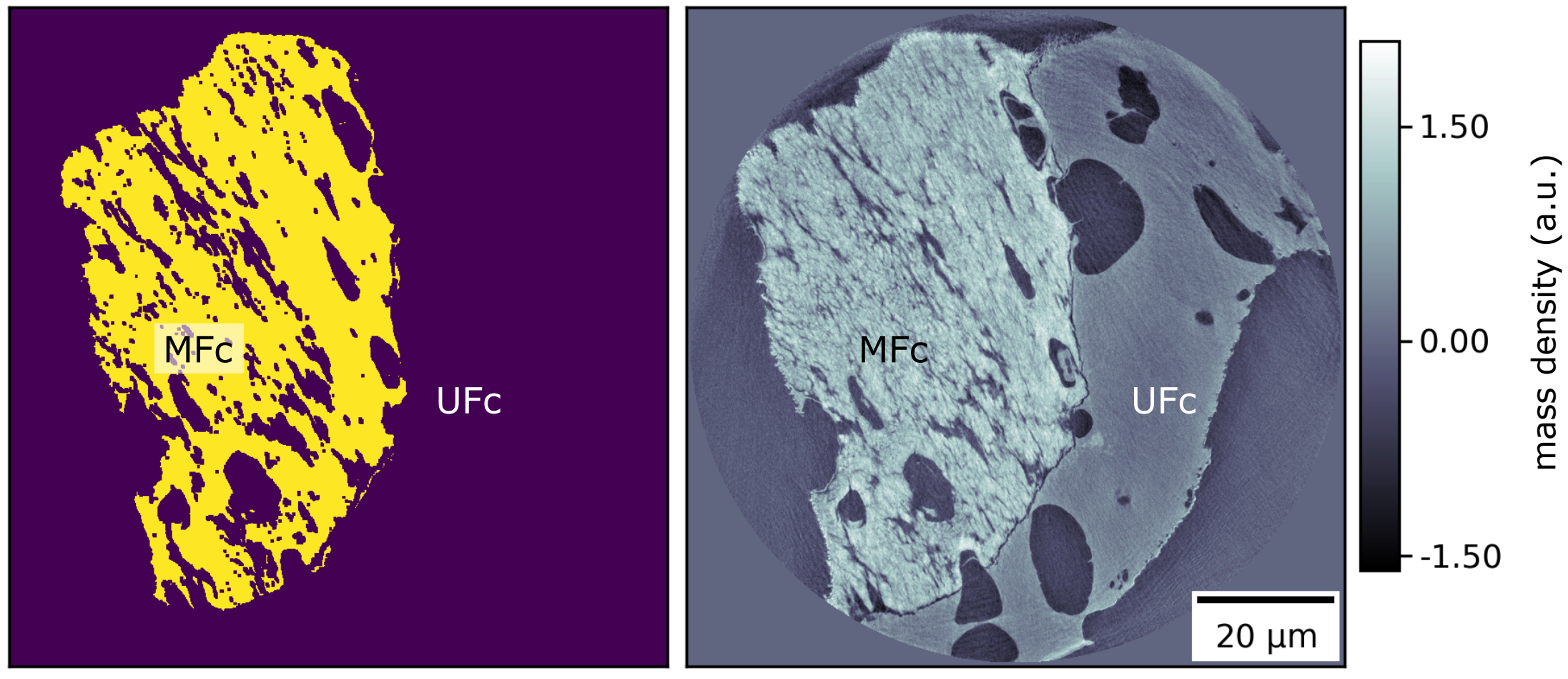

*SI Figure 12* ***Segmentation of holotomo data*** *Mass density shown on the right and the local iterative histogram thresholding in 3D yields the mineralized fibrocartilage tissue*